\documentclass[a4paper,fleqn]{cas-sc}

\usepackage[numbers,sort&compress]{natbib}

\def\tsc#1{\csdef{#1}{\textsc{\lowercase{#1}}\xspace}}
\tsc{WGM}
\tsc{QE}

\usepackage{float}
 
\usepackage[section]{placeins}
\begin{document}

\let\WriteBookmarks\relax
\def\floatpagepagefraction{1}
\def\textpagefraction{.001}

\shorttitle{}    

\shortauthors{}  

\title [mode = title]{Thermodynamically Consistent Machine Learning for Fracture Prediction in Heterogeneous Materials}  


\author[1]{Muhamad Alim}

\author[2]{Xiaolong He}

\author[3]{Mei-Na Tsao}
\author[4]{Chi-Hua Yu}

\author[1,5]{J. S. Chen}
\cormark[1]
\ead{jsc137@ucsd.edu}

\affiliation[1]{
    organization={Department of Mechanical and Aerospace Engineering, University of California San Diego},
    city={La Jolla},
    state={CA},
    postcode={92093},
    country={USA}
}

\affiliation[2]{
    organization={Computational and Multiscale Mechanics Group, Synopsys Inc.},
    city={Livermore},
    state={CA},
    postcode={94551},
    country={USA}
}
\affiliation[3]{
    organization={Department of Civil Engineering, National Cheng Kung University},
    city={Tainan City},
    postcode={701},
    country={Taiwan}
}
\affiliation[4]{
    organization={Department of Engineering Science, National Cheng Kung University},
    city={Tainan City},
    postcode={701},
    country={Taiwan}
}
\affiliation[5]{
    organization={Department of Structural Engineering, University of California San Diego},
    city={La Jolla},
    state={CA},
    postcode={92093},
    country={USA}
}

\cortext[1]{Corresponding author}



\begin{abstract}
Understanding fracture behavior is critical for assessing the structural integrity and longevity. Highly time-consuming physics-based fracture simulations motivate the development of machine-learning-based fracture evolution prediction with surrogate models. However, most existing machine-learning models for fracture prediction operate as black boxes and require large training datasets to generalize reliably. In this work, Thermodynamically Consistent Latent Dynamics Identification (tLaSDI) for parametric systems is introduced to predict fracture patterns and evolution in heterogeneous materials. The model learns the thermodynamic state variable in a lower-dimensional latent space using parametric GENERIC Formalism-Informed Neural Networks (pGFINNs) that enforce key thermodynamic principles, such as total energy conservation and entropy generation. This model embeds physics into the neural network architecture using a hard-constraint method that inherently complies with the underlying thermodynamic principles. The proposed tLaSDI-pGFINNs is further enhanced by a Geometry-Conditioned Autoencoder in learning the geometry configuration for effective geometry-guided learning. The energy and entropy evolution histories demonstrate that the model preserves the thermodynamic constraints on unseen testing samples. The results further show that tLaSDI-pGFINNs can more accurately capture both fracture patterns and the subsequent fracture evolution than the purely data-driven Autoencoder Long Short-Term Memory (AE-LSTM) model.

\end{abstract}





\maketitle

\section{Introduction}
Heterogeneous materials with discontinuous material properties pose considerable challenges for fracture modelling due to the roughness of the deformation field and the highly configuration‑dependent nature of fracture evolution within the heterogeneous domain. Mesh-based methods, such as the finite element method (FEM), require tedious re-meshing techniques to model evolving discontinuities, resulting in high computational cost with low accuracy \cite{Ted2023}\cite{Daxini2014}\cite{Rabczuk2006}. The extended finite element method (XFEM) with embedded discontinuities remains ineffective in modelling three-dimensional and complex fracture patterns \cite{Belytschko2009ARO}. A cohesive‑zone model (CZM) in the finite element setting is implemented by inserting cohesive elements or cohesive interfaces between bulk elements, leading to a mesh-dependent solution with respect to element orientation and discretization resolution
\cite{Wu2015}\cite{CHENG2024}. A phase‑field fracture model implemented with the finite element method (FEM) offers a variational formulation without tracking discontinuities explicitly, but often requires excessive mesh resolution near the crack zones for sufficient accuracy \cite{BOURDIN2000}\cite{Ambati2015}\cite{Cervera2022}\cite{zhangG2023}.

Recently, machine learning and neural networks have been used as surrogate models for fracture prediction. Many supervised ML models, such as random forest, support vector machines, and linear regression, have been proposed in predicting fracture properties such as toughness \cite{BagherShemirani2024}. It has been shown that the traditional regression methods fail to capture complex nonlinear fracture mechanics behavior \cite{Li2025}. The Recurrent Neural Network, on the other hand, performed better in the unseen samples. In \cite{Gao2023}, while the Convolutional Neural Network (CNN) was used to predict the damage field, and a scalar regression task is performed to predict the final peak load, no damage evolution was predicted in the composite specimen. In recent years, advanced machine learning algorithms have been employed to generate surrogate-based models in solid mechanics, these include the geometric deep learning for computational solid mechanics \cite{VLASSIS2020}, the physics-constrained data-driven approach for nonlinear solid \cite{QHe,he2020physics,He2021deep}, the thermodynamically consistent machine-learned internal state variable approach for data-driven modeling of path-dependent materials \cite{He2022}, the Sobolev training of thermodynamics-informed neural networks for interpretable elasto-plasticity models with level set hardening \cite{Vlassis2021}\cite{Wang2018}, the manifold learning based data-driven modeling for soft biological tissues \cite{He2021}, the mechanistic deep material networks embedding micro-mechanical principles for multiscale materials modeling \cite{liu2019deep,wei2023ls,he2026systematic}, among others. Alternatively, machine learning algorithms based on energy principles have been developed to enhance computational methods for effective modelling of problems with local features. For example, a neural network-enhanced reproducing kernel particle method have been proposed for modeling strain localization based on potential energy minimization \cite{Baek2022}. Similar neural network enrichment for fracture modelling have been also developed based on thermodynamically consistent potential energy accounting for energy dissipation in the fracture processes \cite{Baek2024}. The energy minimization approach has been extended to model problems with local features without tedious adaptive refinement in the conventional computational methods \cite{Baek2024NadaptiveRitz} and the formulation is applicable to dynamics problems using the principle of least action \cite{Wang2026}. Other machine learning approaches for fracture prediction include CNN-based surrogate for the phase field damage model \cite{Gao2023}, deep learning to predict fracture patterns in crystalline solids \cite{Hsu2020}, prediction of fracture toughness of concrete using the machine learning approach \cite{BagherShemirani2024}, and a comprehensive study of machine learning methods for predicting fractures have been documented in \cite{Li2025}. 

Physics-Informed Neural Networks (PINNs) \cite{Raissi2019} integrate physics into learning, where the residuals of the governing equations are incorporated into the loss function optimization \cite{Lin2026}\cite{enrui2022}. Wang et al. \cite{Wang2026} developed a surrogate model for composite mechanics using PINNs, and showed that PINNs can achieve good accuracy and faster fitting times with smaller sample sizes than the conventional NNs. Razavi et al. \cite{Razavi2026} used Graph Convolutional Neural Networks combined with Long Short-Term Memory (LSTM) architecture to forecast microstructure evolution of Bismuth-Antimony alloys trained using phase-field generated datasets. However, PINNs-related works with soft constraints on governing equations in loss functions encountered the difficulty of optimizing the loss landscape, especially for complex problems, as reported in \cite{Krishnapriyan2021}. Alternatively, approaches to implement physics directly into a neural network have been investigated. The hard-constraint method is an approach where the physics information is embedded into the neural network architecture. Mohan et al. \cite{Mohan2020} proposed a physics Convolutional Autoencoder (PhyCAE) to enforce the incompressibility of a fluid flow within the CNN architecture. A mathematical strategy was implemented in which the network is designed to predict a velocity potential field rather than the velocity directly, and physics is guaranteed by construction. Another notable framework is the General Equation for Non-Equilibrium Reversible-Irreversible Coupling (GENERIC) \cite{Grmela1997, ttinger2018}, which provides a thermodynamic formalism for modeling the evolution of systems operating beyond equilibrium. GENERIC has been applied to discover the thermodynamic structure in a constitutive model of soft materials from experimental measurements \cite{Ghnatios2019} and identify the missing dissipative correction from the pseudo-experimental data obtained from a Mooney-Rivlin hyperelastic model \cite{Gonzlez2019}. Due to the effectiveness of GENERIC to discover the underlying thermodynamics, this mathematical framework has been expanded into a machine learning application. Lee et al. \cite{Lee2021} investigated the implementation of the GENERIC formalism into neural networks for a damped nonlinear oscillator. They compared a penalty-based model, in which the GENERIC structure is enforced softly through additional loss terms, with a hard-constrained approach, termed the GENERIC Neural ODE (GNODE), where the GENERIC structure is embedded directly into the neural network architecture through a bracket-based parameterization.  Their results demonstrated that GNODE significantly outperforms both the penalty-based model and the black-box NODEs in predicting entropy and mechanical energy. In particular, GNODE preserves the thermodynamic structure by construction, ensuring non-decreasing entropy with consistent energy dissipation, and achieves accurate predictions in both interpolation and extrapolation regimes. In contrast, the penalty-based model violates the second law of thermodynamics by allowing negative entropy production, indicating that soft enforcement of the thermodynamics constraints is insufficient to guarantee physical consistency. Although the penalty-based approach performs well within the training (interpolation) range, it fails to generalize in the extrapolation regime due to the breakdown of thermodynamic conditions. Furthermore, Zhang et al. \cite{Zhang2022GFINN} proposed the GENERIC Formalism Informed Neural Networks (GFINNs), a generalized framework that decomposes the GENERIC dynamics into separate learnable energy, entropy, reversible, and dissipative components. This formulation enables the incorporation of varying levels of prior physical knowledge while preserving the thermodynamic structure by construction.

The above-mentioned GFINNs, however, learn the state variables in a physical space, which limits their application to learning high-dimensional physical systems. To address this, Park et al. \cite{Park2024} proposed a thermodynamics-informed latent space dynamics identification (tLaSDI), a non-intrusive reduced-order model (ROM) that represents a high-dimensional solution in a low-dimensional space. This model learns the thermodynamic structure by incorporating GFINNs in the latent space. A parametric partial differential equation (PDE) was also investigated \cite{Park2024}, where tLaSDI was applied to predict the solution of the 1D Burgers’ equation.  Furthermore, He et al. \cite{He2025} extends the tLaSDI framework by introducing parametric GFINNs (pGFINNs) to replace  the hypernetwork-based architecture, enabling efficient learning of parametric latent dynamics. However, the application of these thermodynamics-informed latent-space frameworks to predict fracture evolution in heterogeneous materials remains limited.

This work develops a thermodynamics-informed, geometry-conditioned latent space dynamics identification framework for accelerated, geometry-aware fracture modeling in heterogeneous materials. The proposed framework integrates geometry-conditioned convolutional autoencoders for nonlinear dimensionality reduction with a pGFINN for thermodynamically consistent fracture evolution in the latent space. To demonstrate its effectiveness, the framework is applied to a range of fracture problems involving heterogeneous materials and diverse parameterizations, including variations in boundary conditions, inclusion geometries and spatial distributions, as well as crack initiation locations. 

The paper is organized as follows. Section 2 introduces the GENERIC Formalism and the Parametric GENERIC Formalism Informed Neural Networks for general mechanics problems. Section 3 discusses the tLaSDI-pGFINNs formulation and Geometry Conditioned Autoencoder for fracture problems. The Dataset generation and numerical setup are introduced in Section 4. Section 5 demonstrates the effectiveness of the proposed approach by analyzing three numerical examples with varying levels of complexity to investigate the performance of the proposed model: (1) a single-cavity problem with varying loading conditions,
(2) a single-inclusion problem with varying inclusion locations and diameters, and (3) a multiple-inclusion problem with varying crack initiation locations. The concluding remarks are summarized in Section 6.

\section{Thermodynamics-Informed Learning Framework}
\subsection{GENERIC Formalism}
General Equation for Non-Equilibrium Reversible and Irreversible Coupling (GENERIC) \cite{Grmela1997}, also known as the metriplectic formalism \cite{morrison1986paradigm}, is a mathematical framework that describes a system beyond thermodynamic equilibrium and can be applied to dynamic systems that evolve over time. Proposed by Grmela et al.  \cite{Grmela1997} as a systematic framework for describing non-Newtonian fluid behavior, this formalism ensures that the model built upon it satisfies the thermodynamic conditions without any prior assumptions. GENERIC time rate of a state variable $\mathbf{z}$ is described as follows

\begin{equation}
\label{eq:generic}
\dot{\mathbf{z}}
=
\mathbf{L}(\mathbf{z}) \cdot
\boldsymbol{\nabla}_{\mathbf{z}} E(\mathbf{z})
+
\mathbf{M}(\mathbf{z}) \cdot
\boldsymbol{\nabla}_{\mathbf{z}} S(\mathbf{z}),
\end{equation}

subject to

\begingroup
\setlength{\abovedisplayskip}{0pt}
\setlength{\belowdisplayskip}{0pt}

\begin{equation}
\label{eq:d1}
\mathbf{L}(\mathbf{z})\!\cdot\!\boldsymbol{\nabla}_{\mathbf{z}} S(\mathbf{z}) = \mathbf{0},
\end{equation}

\begin{equation}
\label{eq:d2}
\mathbf{M}(\mathbf{z})\!\cdot\!\boldsymbol{\nabla}_{\mathbf{z}} E(\mathbf{z}) = \mathbf{0},
\end{equation}

\begin{equation}
\label{eq:s1}
\mathbf{L}(\mathbf{z}) = -\mathbf{L}^{T}(\mathbf{z}),
\end{equation}

\begin{equation}
\label{eq:s2}
\mathbf{M}(\mathbf{z}) = \mathbf{M}^{T}(\mathbf{z}),
\quad
\mathbf{x}\!\cdot\mathbf{M}(\mathbf{z})\!\cdot\mathbf{x}\ge 0,
\quad
\forall\,\mathbf{x} \neq \mathbf{0},
\end{equation}
\endgroup
\\
where $\boldsymbol{\nabla}_{\mathbf{z}}(\cdot) = \partial(\cdot)/\partial \mathbf{z}$. The first two terms on the right hand side of Eq.~\eqref{eq:generic} account for the reversible and non-dissipative processes, where $E$ is the energy of the system. In classical mechanics, it corresponds to the Hamiltonian dynamics describing the evolution of position and momentum. The matrix operator $\mathbf{L}(\mathbf{z})$, called the Poisson matrix with an anti-symmetric property, is multiplied to this reversible term. The third and forth terms on the right hand side of Eq.~\eqref{eq:generic} account for the remaining processes in a system that corresponds to the dissipative phenomena, where $S$ is the entropy of the system, and this irreversible term is associated with a symmetric positive semi-definite Friction matrix $\mathbf{M}(\mathbf{z})$.
The vector $\mathbf{z}$ denotes the state variables representing the independent physical states of the system. For example, in a study conducted by \cite{Ghan2019}, state variables are the position (generalized coordinate) and momentum for a simple pendulum case, whereas for a thermoelastic pendulum problem, the state variables include the mechanical motion, thermal effects (e.g., entropy, heat transfer) and the dissipation. Satisfaction of Eqs.~\eqref{eq:d1} and \eqref{eq:d2}, called the degeneracy conditions, ensure the absence of contribution from the energy potential to the entropy production and likewise, entropy production does not affect the reversible dynamics. Using the above conditions shown in Eq.~\eqref{eq:d1}-\eqref{eq:s2}, the first law of thermodynamics which is the energy conversation 
\begin{equation}
\label{conserv}
\dot{E}
=
\boldsymbol{\nabla}_{\mathbf{z}} E
\!\cdot\!
\dot{\mathbf{z}}
=
\boldsymbol{\nabla}_{\mathbf{z}} E \!\cdot\!
\left(
\mathbf{L} \cdot
\boldsymbol{\nabla}_{\mathbf{z}} E
+
\mathbf{M} \cdot
\boldsymbol{\nabla}_{\mathbf{z}} S
\right)
=0,
\end{equation}
due to the conditions in Eqs. \eqref{eq:d2} and \eqref{eq:s1}. Similarly, the change of entropy is guaranteed to be non-decreasing over time, which corresponds to the second law of thermodynamics
\begin{equation}
\label{diss}
\dot{S}
=
\boldsymbol{\nabla}_{\mathbf{z}} S 
\!\cdot\!
\dot{\mathbf{z}}
=
\boldsymbol{\nabla}_{\mathbf{z}} S \!\cdot\!
\left(
\mathbf{L} \cdot
\boldsymbol{\nabla}_{\mathbf{z}} E
+
\mathbf{M} \cdot
\boldsymbol{\nabla}_{\mathbf{z}} S
\right)
\ge 0,
\end{equation}
where the conditions in \eqref{eq:d1} and \eqref{eq:s2} have been used. 
From a geometric perspective, the trajectory of $\mathbf{z}$ remains tangent to the level curves of $E$ and transverse to those of $S$, reflecting the fact that GENERIC/metriplectic dynamics are a combination of Hamiltonian and generalized gradient flows.
When $\mathbf{M}=\mathbf{0}$, $E=H$ corresponds to the Hailtonian function and $\mathbf{L}=\tilde{\mathbf{J}}=\sqrt{-\mathbf{I}}$, where $\tilde{\mathbf{J}}$ denotes the standard symplectic matrix and $\mathbf{I}$ is an identity matrix, and Eq.\eqref{eq:generic} reduces to Hamilton's equations of motion $\dot{\mathbf{z}} = \tilde{\mathbf{J}} \!\cdot\!\boldsymbol{\nabla}_{\mathbf{z}} H$. 
Similarly, when $\mathbf{L}=\mathbf{0}$ and $S=-G$, where $G$ is some scalar functional (typically a Lyapunov or free-energy function), Eq.\eqref{eq:generic} reduces to a generalized gradient flow $\dot{\mathbf{z}} = -\mathbf{M}\!\cdot\!\boldsymbol{\nabla}_{\mathbf{z}} G$.

The GENERIC/metriplectic forms of various physical systems have been studied in the literature, including the compressible Navier-Stokes equations \cite{morrison2024some}, the Smoluchowski equation for colloidal suspensions \cite{wagner2001smoluchowski}, polymer melts and solutions \cite{ait1999volume}, 
dissipative magnetohydrodynamics \cite{materassi2012metriplectic}, rigid-body dynamics \cite{materassi2018metriplectic}, Hamiltonian systems with friction \cite{caligan2016conservative}, the Vlasov–Fokker–Planck equation \cite{duong2013generic}, among others.

\subsection{Parametric Generic Formalism Informed Neural Networks}
Due to the general description of thermodynamic properties, structure-preserving model reduction and dynamics identification of GENERIC/metriplectic systems have become an active area of research \cite{hernandez2021structure,Lee2021,Zhang2022GFINN,gruber2023energetically,gruber2023reversible}. A GENERIC formalism-informed neural networks (GFINNs) proposed by \citet{Zhang2022GFINN} can yield an exact satisfaction of the symmetrices and degeneracy conditions in Eqs.\eqref{eq:d1}-\eqref{eq:s2} through structure-informed neural network architectures to enable thermodynamically consistent data-driven learning of physical processes. However, GFINNs do not inherently support differential equations with parametric variability in their governing operators. To address this limitation, He et al. \cite{He2025} introduced a parametric GFINN (pGFINN) framework to extend the method's capability for modeling parametric dynamical systems. 

Figure~\ref{fig:Figure5} illustrates the architecture of pGFINNs, where the GFINN networks are parameterized by $\boldsymbol{\mu}$. A key operation in the pGFINNs is the construction of the operator matrices that automatically satisfy the symmetry properties and degeneracy conditions in Eqs.\eqref{eq:d1}-\eqref{eq:s2}. To this end, the Poisson matrix $\mathbf{L}$ and the Friction matrix $\mathbf{M}$ are designed to ensure $\boldsymbol{\nabla}_{\mathbf{z}}S \in \text{ker} (\mathbf{L})$ and $\boldsymbol{\nabla}_{\mathbf{z}}E \in \text{ker} (\mathbf{M})$, respectively.
Building on matrix factorization, the Poisson matrix $\mathbf{L}$ is represented by
\begin{equation}\label{eq:L_factorization}
    \mathbf{L} = \mathbf{Q}_S\!\cdot\!\mathbf{B}_L\!\cdot\!\mathbf{Q}_S,
\end{equation}
where $\mathbf{B}_L=\mathbf{T}_L^{T}-\mathbf{T}_L$ is a skew-symmetric matrix, with $\mathbf{T}_L$ as a learnable triangular matrix.
To enforce the entropy degeneracy condition, $\mathbf{L}\!\cdot\!\boldsymbol{\nabla}_{\mathbf{z}}S=\mathbf{0}$, a matrix-valued function $\mathbf{Q}_S$ is constructed as follows. 
As illustrated in Figure~\ref{fig:Figure5}(a), the state variable $\mathbf{z}$ is first passed into an NN to extract the entropy $S$. Then, the corresponding gradient, $\boldsymbol{\nabla}_{\mathbf{z}}S$, is computed and used to construct $\mathbf{Q}_S$, whose $j$-th row is defined as $(\mathbf{W}_j\cdot\!\boldsymbol{\nabla}_{\mathbf{z}}S)^T$, where $\mathbf{W}_j$, $j=1,...,K$ are learnable skew-symmetric matrix fields. 
By construction, this formulation guarantees that $\mathbf{Q}_S\cdot\!\boldsymbol{\nabla}_{\mathbf{z}}S=\mathbf{0}$, thereby ensuring that the entropy degeneracy condition $\mathbf{L}\!\cdot\!\boldsymbol{\nabla}_{\mathbf{z}}S=\mathbf{0}$ is satisfied.
This follows from the property of skew-symmetric matrices, i.e., $\mathbf{x}\!\cdot\!\mathbf{W}_j\cdot\!\mathbf{x}=0$, $\forall \mathbf{x} \neq \mathbf{0}$.
The computational graph for constructing the Poisson matrix $\mathbf{L}$ is depicted in Figure~\ref{fig:Figure5}(a). 

Similarly, the Friction matrix $\mathbf{M}$ is expressed as
\begin{equation}\label{eq:M_factorization}
    \mathbf{M} = \mathbf{Q}_E\!\cdot\!\mathbf{B}_M\!\cdot\!\mathbf{Q}_E,
\end{equation}
where $\mathbf{B}_M=\mathbf{T}_M^{T}\mathbf{T}_M$ is a symmetric matrix, with $\mathbf{T}_M$ as a learnable triangular matrix.
To enforce the energy degeneracy condition $\mathbf{M}\!\cdot\!\boldsymbol{\nabla}_{\mathbf{z}}E=0$, a matrix-valued function $\mathbf{Q}_E$ is constructed, whose $j$-th row is defined as $(\mathbf{W}_j\cdot\!\boldsymbol{\nabla}_{\mathbf{z}}E)^T$. By design, this ensures that $\mathbf{Q}_E\cdot\!\boldsymbol{\nabla}_{\mathbf{z}}E=\mathbf{0}$, thereby satisfying the energy degeneracy condition $\mathbf{M}\cdot\!\boldsymbol{\nabla}_{\mathbf{z}}E=\mathbf{0}$.
The computational graph for constructing the Friction matrix $\mathbf{M}$ is illustrated in Figure~\ref{fig:Figure5}(b). 

To account for parametric dependence, the networks associated with $S$, $E$, $\mathbf{B}_L$, and $\mathbf{B}_M$ are parameterized by $\boldsymbol{\mu}$. This enables the pGFINN framework to effectively represent parameter-dependent dynamics while preserving free energy conservation and ensuring entropy production across the entire parameter space.

\begin{center}
\refstepcounter{figure}

\includegraphics[width=0.95\textwidth]{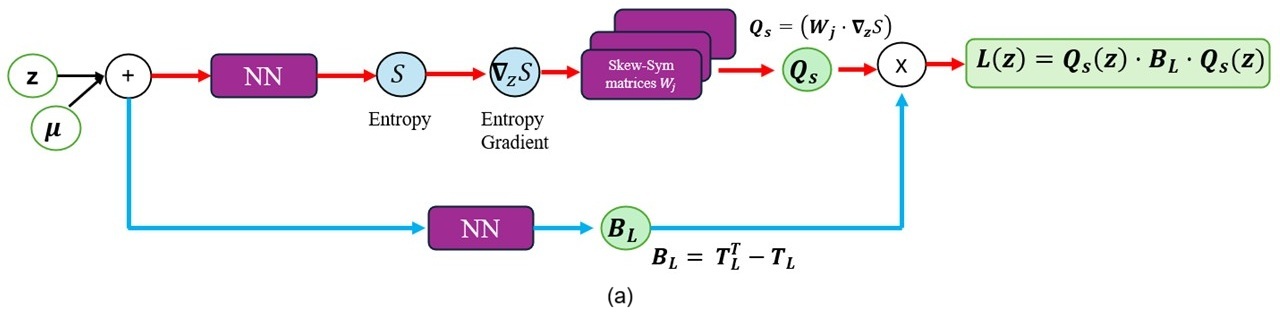}

\vspace{0.5em}

\includegraphics[width=0.95\textwidth]{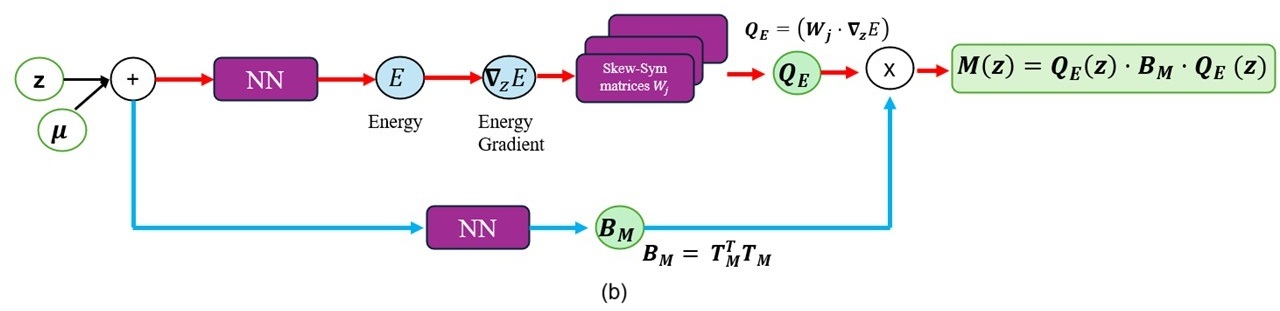}
\\[0.5em]
\textbf{Figure \thefigure.}
(a) Networks for entropy and the Poisson matrix $\mathbf{L}$.
(b) Networks for energy and the friction matrix $\mathbf{M}$.

\label{fig:Figure5}
\end{center}
\FloatBarrier
\section{Thermodynamics-Based Latent Dynamics Framework for Fracture Prediction}
A parametric dynamical system governed by a system of ordinary differential equations (ODEs) is considered
\begin{equation}\label{eq.govern_eqn}
    \dot{\mathbf{u}}(t; \boldsymbol{\mu}) = \mathbf{f}(\mathbf{u},t; \boldsymbol{\mu}), \quad t \in [0,T], \quad \mathbf{u}(0; \boldsymbol{\mu}) = \mathbf{u}_0(\boldsymbol{\mu}) \in \mathbb{R}^{N_u},
\end{equation}
where $\mathbf{u}(t; \boldsymbol{\mu})$ denotes the time-dependent state vector parametrized by $\boldsymbol{\mu}$. 
The initial condition $\mathbf{u}_0$ depends on the parameter $\boldsymbol{\mu} \in \mathcal{D} \subseteq \mathbb{R}^{N_{\mu}}$. where $\mathcal{D}$ is the admissible parameter domain.
$\mathbf{f}$ is a sufficiently regular vector field ensuring existence and uniqueness of the solution, and 
$T \in \mathbb{R}_{+}$ denotes the final time.
Eq. \eqref{eq.govern_eqn} can be considered as a semi-discrete form of an underlying system of PDEs. 
Its numerical solution typically requires explicit or implicit time integration schemes, which can become computationally expensive for large-scale systems with high-dimensional state vectors, i.e., large $N_u$. 
In this work, we aim to develop a thermodynamically consistent ROM framework for efficient and accurate fracture prediction in heterogeneous materials.

\subsection{tLaSDI-pGFINN Formulation}\label{sec:loss}
To model high-dimensional dynamical systems while preserving thermodynamic principles, Park et al. \cite{Park2024} proposed thermodynamics-informed latent space dynamics identification (tLaSDI), a non-intrusive ROM that combines an autoencoder for nonlinear dimensionality reduction with GFINN for latent dynamics prediction.
To further enable efficient learning of parametric latent dynamics, He et al. \cite{He2025} extended this framework by integrating parametric GFINN, resulting in the tLaSDI-pGFINN.
Figure~\ref{fig:tLaSDIpGFINN_graph} illustrates the inference mode of tLaSDI-pGFINN. The high-dimensional initial state $\mathbf{u}_0$ is encoded into a low-dimensional latent representation $\mathbf{z}_0$ using the convolutional encoder $\boldsymbol{\phi}_e$. Given $\mathbf{z}_0$ and the parameter $\boldsymbol{\mu}$, the pGFINN predicts thermodynamically consistent latent states at subsequent time step. The dashed loop indicates the autoregressive prediction process, in which the predicted latent state is recursively fed back into the pGFINN to generate future latent states $\mathbf{z}_+$. 
After all latent states are obtained, the convolutional decoder $\boldsymbol{\phi}_d$ reconstructs the corresponding high-dimensional states in the physical space.

\begin{center}
\refstepcounter{figure}
\includegraphics[width=0.9\textwidth]{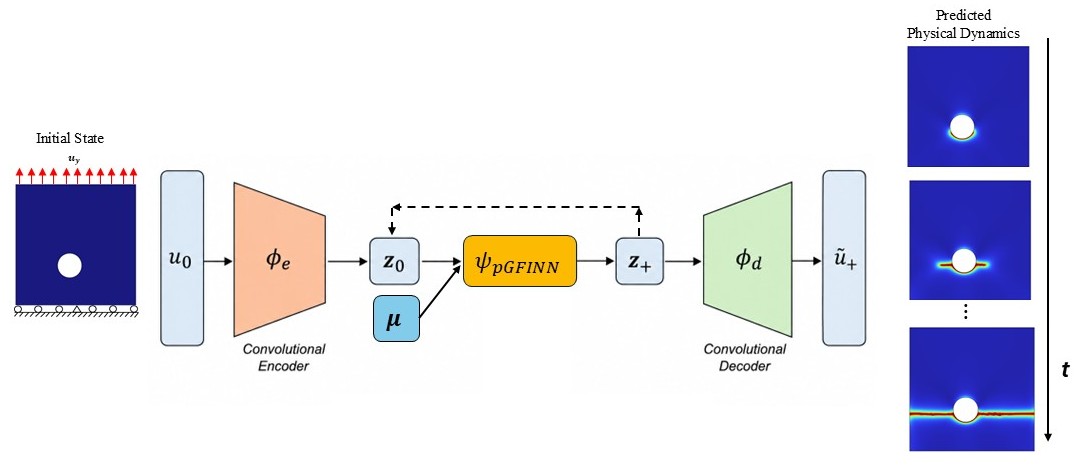}
\\[0.0em]
\textbf{Figure~\thefigure:} Schematic of tLaSDI-pGFINN in inference mode.
\label{fig:tLaSDIpGFINN_graph}
\end{center}

The loss function of tLaSDI-pGFINN consists of four terms
\begin{equation}
\mathcal{L}(\boldsymbol{\theta})
=
\mathcal{L}_{int}(\boldsymbol{\theta})
+
\lambda_{rec}\mathcal{L}_{rec}(\boldsymbol{\theta})
+
\lambda_{jac}\mathcal{L}_{jac}(\boldsymbol{\theta})
+
\lambda_{mod}\mathcal{L}_{mod}(\boldsymbol{\theta}),
\label{eq:totalloss}
\end{equation}
where $\boldsymbol{\theta}$ denotes the collection of all trainable parameters.
The integration loss, $\mathcal{L}_{int}$, measures the inconsistency between the encoder-predicted latent state at the next time step and the corresponding state obtained by numerically integrating the learned latent dynamics over the time interval.
\begin{equation}\label{eq.loss_ini}
\mathcal{L}_{int}
=
\sum_{\boldsymbol{\mu}}\sum_{n}
\left\|
\boldsymbol{\phi}_e(\mathbf{u}_{n+1})
-
\boldsymbol{\phi}_e(\mathbf{u}_n)
-
\int_{t_n}^{t_{n+1}}
\boldsymbol{\psi}_{\text{pGFINN}}
(\mathbf{z}(t),\mu,t)\,dt
\right\|^2.
\end{equation}
Here, $\mathbf{u}_n$ denotes the full-order physical state at $t_n$; $\boldsymbol{\phi}_e$ is the encoder mapping from the physical space to the latent space; $\boldsymbol{\psi}_{\text{pGFINN}}$ represents the latent dynamics function learned by the pGFINN; $\mathbf{z}(t)$ is the latent variable; $\boldsymbol{\mu}$ denotes the parameter of the dynamical system.
In other words, the integration loss enforces the following approximate equality
\begin{equation}\label{eq.loss_ini_1}
    \boldsymbol{\phi}_{\text{e}}(\mathbf{u}_{n+1}) \approx
    \boldsymbol{\phi}_{\text{e}}(\mathbf{u}_{n}) + 
    \int_{t_n}^{t_{n+1}} \boldsymbol{\psi}_{\text{pGFINN}}\big(\boldsymbol{z}(t),\boldsymbol{\mu},t\big) dt,
\end{equation}
providing a consistency condition between pGFINN integration and the encoder dynamics.


Then, the reconstruction loss, $\mathcal{L}_{rec}$, measures how closely the autoencoder's output matches the original input. This loss quantifies the information loss during compression-decompression process. The reconstruction loss is defined as 

\begin{equation}
\mathcal{L}_{rec}
=
\sum_{\boldsymbol{\mu}}\sum_n
\left\|
\mathbf{u}_n
-
\boldsymbol{\phi}_{d}
\!\left(
\boldsymbol{\phi}_{e}(\mathbf{u}_n)
\right)
\right\|^2,
\label{eq:loss_rec}
\end{equation}
where $\mathbf{\phi}_d$ defines the decoder mapping from latent space to physical space.

The Jacobian loss, $\mathcal{L}_{jac}$, is introduced to enhance the model’s ability to preserve the underlying dynamics of the system. This loss is derived from ROM error analysis presented in \ref{sec:error_analysis}, which shows the approximation error does not only rely on the reconstruction output but also the accuracy of the model to capture the derivatives or dynamics of the full system \cite{Park2024}. Therefore, this loss is included to reduce an upper bound of the ROM approximation error. The Jacobian loss is defined as
\begin{equation}\label{eq.loss_jac_2}
\mathcal{L}_{jac}
=
\sum_{\boldsymbol{\mu}}\sum_{n}
\left\|
\left(
\mathbf{I}
-
\mathbf{J}(\mathbf{u}_n)
\right)
\dot{\mathbf{u}}_n
\right\|^2,
\end{equation}
where $\dot{\mathbf{u}}_n$ denotes the time derivative of the physical state vector, and $\mathbf{I}$ is the identity matrix. 
The operator $\mathbf{J}(\cdot) = \mathbf{J}_{d}\big(\boldsymbol{\phi}_{e} (\cdot)\big) \mathbf{J}_{e} (\cdot)$ represents the Jacobian of the autoencoder.
It is obtained via the chain rule as the composition of the decoder and encoder Jacobians, denoted by 
$\mathbf{J}_{d}(\cdot)$ and $\mathbf{J}_{e}(\cdot)$, respectively.
It is worth noting that this corresponds to a Sobolev-type loss that penalizes infinitesimal inconsistency between the state $\mathbf{u}_n$ and its autoencoder reconstruction, enforcing
\begin{equation}\label{eq.loss_jac_1}
    \dot{\mathbf{u}}_{n} \approx \frac{d}{dt} \boldsymbol{\phi}_{d} \big( \boldsymbol{\phi}_{e} (\mathbf{u}_{n}) \big) =
    \mathbf{J}_{d} \big(\boldsymbol{\phi}_{e} (\mathbf{u}_{n})\big) 
    \mathbf{J}_{e} (\mathbf{u}_{n}) \dot{\mathbf{u}}_{n} = 
    \mathbf{J}(\mathbf{u}_{n}) \dot{\mathbf{u}}_{n}.
\end{equation}
In the absence of the derivative data $\dot{\mathbf{u}}_{n}$, the Jacobian loss is alternatively defined as
\begin{equation}\label{eq.loss_jac_3}
    \mathcal{L}_{Jac} = 
    \sum_{\boldsymbol{\mu}} \sum_{n}
    \Big\lVert 
        \mathbf{I} - \mathbf{J}(\mathbf{u}_{n})
    \Big\rVert_{F}^2,
\end{equation}
where $\lVert \cdot \rVert_F$ denotes the Frobenius norm. 
Since the latent space has a much lower dimension than the full-order physical space, the autoencoder Jacobian cannot be full rank, implying $\mathbf{J}(\cdot)\neq\mathbf{I}$. This structural limitation motivates the introduction of the final modeling loss $\mathcal{L}_{mod}$.
In particular, by adding and substracting a suitable term and applying the Cauchy-Schwarz inequality, one obtains
\begin{equation}\label{eq.loss_mod_1}
    \lVert \big(\mathbf{I} - \mathbf{J}(\mathbf{u}_{n}) \big) \dot{\mathbf{u}}_{n} \rVert 
    \le
    \lVert \dot{\mathbf{u}}_{n} - \mathbf{J}_{d}\big(\boldsymbol{\phi}_{\text{e}} (\mathbf{u}_{n})\big) \boldsymbol{\psi}_{\text{pGFINN}} \big( \boldsymbol{\phi}_{e} (\mathbf{u}_{n}) \big) \rVert
    +
    \lVert \mathbf{J}_{d}\big(\boldsymbol{\phi}_{\text{e}} (\mathbf{u}_{n})\big) \rVert \cdot 
    \lVert \mathbf{J}_{e} (\mathbf{u}_{n}) \dot{\mathbf{u}}_{n} - \boldsymbol{\psi}_{\text{pGFINN}} \big( \boldsymbol{\phi}_{e} (\mathbf{u}_{n}) \big)\rVert.
\end{equation}
Accordingly, the modeling loss $\mathcal{L}_{mod}$ is defined as
\begin{equation}\label{eq.loss_mod_2}
\mathcal{L}_{mod}
=
\sum_{\boldsymbol{\mu}}\sum_{n}
\left\|
\dot{\mathbf{u}}_n
-
\mathbf{J}_d(\boldsymbol{\phi}_e(\mathbf{u}_n))
\boldsymbol{\psi}_{\text{pGFINN}}(\boldsymbol{\phi}_e(\mathbf{u}_n))
\right\|^2
+
\left\|
\mathbf{J}_e(\mathbf{u}_n)\dot{\mathbf{u}}_n
-
\boldsymbol{\psi}_{\text{pGFINN}}(\boldsymbol{\phi}_e(\mathbf{u}_n))
\right\|^2,
\end{equation}
where the first term on the right-hand side quantifies the modeling error in the full-order dynamics, obtained by projecting the latent dynamics predicted by pGFINN through the decoder.
The second term measures the discrepancy between the encoder-derived latent velocity field, $\mathbf{J}_{e} (\mathbf{u}_{n}) \dot{\mathbf{u}}_{n}$,
and the pGFINN prediction, $\boldsymbol{\psi}_{\text{pGFINN}} \big( \boldsymbol{\phi}_{e} (\mathbf{u}_{n}) \big)$.

The loss function in Eq.\eqref{eq:totalloss} can be interpreted as an error bound for tLaSDI-pGFINN under the assumptions that the decoder Jacobian is bounded and Lipschitz continuous, and that the pGFINN predictions remain bounded \cite{Park2024,He2025}. 
Importantly, this formulation enables joint training of the autoencoder and the latent dynamics model, thereby enforcing consistency in the learned latent dynamics while simultaneously minimizing modeling errors in the full-order system. Such joint training strategies have been shown to offer advantages in prior works \cite{champion2019data,he2023glasdi,Park2024,bonneville2024gplasdi,he2025physics,He2025}. In particular, explicitly coupling the autoencoder with the latent dynamics model and thus imposing structure on the latent space has been demonstrated to be critical for enhancing both predictive accuracy and generalization performance. Furthermore, an upper bound on the tLaSDI-pGIFNN prediction error, expressed in terms of the loss function Eq.\eqref{eq:totalloss}, is presented in Section \ref{sec:error_analysis}, providing a theoretical justification for this training approach.

It is worth noting that constructing the full tLaSDI-pGFINN loss in Eq. \eqref{eq:totalloss} requires derivative information in the training data, particularly for the Jacobian loss in Eq. \eqref{eq.loss_jac_2} and the modeling loss in Eq. \eqref{eq.loss_mod_2}.
In scenarios where such information is difficult to obtain, such as in the presence of noisy data, the strong form ODE in Eq. \eqref{eq.govern_eqn} can be transformed to a weak form by using compactly supported test functions. This reformulation transfers temporal derivatives from the state variables to the test function, thereby mitigating noise sensitivity and eliminating the need for pointwise derivatives evaluations \citep{messenger2021weak,tran2024weak,he2025physics,bonneville2024comprehensive}.
Moreover, the weak formulation-based GFINN \cite{park2026wgfinns} can be employed to identify thermodynamically consistent latent dynamics under noisy conditions.

\subsection{Error Analysis}\label{sec:error_analysis}
This section provides a bound for the error of the tLaSDI-pGFINN approximation, following an approach similar to \cite{Park2024}, with a proof included for completeness.
For simplicity, parametric dependence is omitted in the analysis. The resulting bound can be interpreted as holding pointwise for each parameter instance $\boldsymbol{\mu}$.
We begin by expressing the tLaSDI-pGFINN approximation error as
\begin{equation}\label{eq.tlasdi_error}
    \mathbf{e}(t) = \mathbf{u}(t) - \Tilde{\mathbf{u}}(t) 
    = \big( \mathbf{u}(t) - \hat{\mathbf{u}}(t) \big)
    + \big( \hat{\mathbf{u}}(t) - \Tilde{\mathbf{u}}(t) \big),
\end{equation}
where $\hat{\mathbf{u}}(t) = \boldsymbol{\phi}_{d}\big( \boldsymbol{\phi}_{e} (\mathbf{u}(t)) \big)$ denotes the the autoencoder reconstruction, and $\Tilde{\mathbf{u}}(t) = \boldsymbol{\phi}_{d}\big(\mathbf{z}(t)\big)$ denotes the tLaSDI-pGFINN approximation.  
Assuming a consistent initialization $\mathbf{z}(t_0) = \boldsymbol{\phi}_{e}\big(\mathbf{u}(t_0)\big)$, the latent dynamics evolve according to $\dot{\mathbf{z}} = \mathbf{f}^r(\mathbf{z})$, where $\mathbf{f}^r$ denotes the pGFINN model. Our objective is to bound the approximation error $\mathbf{e}(t)$ in terms of quantities associated with the loss terms $\mathcal{L}_{int},\mathcal{L}_{rec},\mathcal{L}_{Jac},$ and $\mathcal{L}_{mod}$. 

To this end, we decompose the error into the autoencoder reconstruction error $\mathbf{e}_{\text{AE}}(t) = \mathbf{u}(t) - \hat{\mathbf{u}}(t)$ and dynamics identification error $\mathbf{e}_{\text{DI}}(t) = \hat{\mathbf{u}}(t) - \Tilde{\mathbf{u}}(t)$.
For the reconstruction error, differentiation yields the initial value problem
\begin{equation}\label{eq.AE_error1}
    \dot{\mathbf{e}}_{\text{AE}} 
    = \dot{\mathbf{u}} - \dot{\hat{\mathbf{u}}}
    = (\mathbf{I} - \mathbf{J}(\mathbf{u})) \dot{\mathbf{u}}, 
    \qquad
    \mathbf{e}_{\text{AE}}(t_0) = \mathbf{u}(t_0) - \boldsymbol{\phi}_{d} \big( \boldsymbol{\phi}_{e} (\mathbf{u}(t_0)) \big).
\end{equation}
Integrating over time gives
\begin{equation}\label{eq.AE_error2}
    \mathbf{e}_{\text{AE}} 
    = \mathbf{e}_{\text{AE}}(t_0) 
    + \int_{t_0}^t \big(\mathbf{I} - \mathbf{J}(\mathbf{u}(s)) \big) \dot{\mathbf{u}}(s) ds,
\end{equation}
which can be bounded for any $\alpha\in[0,1]$,
\begin{equation}\label{eq.AE_error_bound}
    \lVert \mathbf{e}_{\text{AE}}(t) \rVert
    \le \alpha \Big( 
    \lVert \mathbf{e}_{\text{AE}}(t_0) \rVert
    + \int_{t_0}^t \big\lVert \big(\mathbf{I} - \mathbf{J}(\mathbf{u}(s)) \big) \dot{\mathbf{u}}(s) \big\rVert ds \Big) 
    + (1-\alpha) \lVert \mathbf{e}_{\text{AE}}(t) \rVert. 
\end{equation}
For the dynamics identification error, we obtain
\begin{equation}\label{eq.DI_error1}
    \dot{\mathbf{e}}_{\text{DI}} 
    = \dot{\hat{\mathbf{u}}} - \dot{\tilde{\mathbf{u}}}
    = \mathbf{J}(\mathbf{u}) \mathbf{f}(\mathbf{u}) - \mathbf{J}_{d}(\mathbf{z}) \mathbf{f}^r(\mathbf{z}), 
    \qquad
    \mathbf{e}_{\text{DI}}(t_0) = \mathbf{0}.
\end{equation}
where $\mathbf{f}(\mathbf{u})$ is the true vector field of the original high-dimensional dynamical system. Adding and subtracting $\mathbf{J}_{d}(\boldsymbol{\phi}_{e}(\mathbf{u})) \mathbf{f}^r(\mathbf{z})$ on the right-hand side yields 
\begin{equation}\label{eq.DI_error2a}
    \mathbf{J}(\mathbf{u}) \mathbf{f}(\mathbf{u}) - \mathbf{J}_{d}(\mathbf{z}) \mathbf{f}^r(\mathbf{z})
    = \Big( \mathbf{J}(\mathbf{u}) \mathbf{f}(\mathbf{u}) 
    - \mathbf{J}_{d}(\boldsymbol{\phi}_{e}(\mathbf{u})) \mathbf{f}^r(\mathbf{z}) \Big)
    + \Big( \mathbf{J}_{d}(\boldsymbol{\phi}_{e}(\mathbf{u})) \mathbf{f}^r(\mathbf{z}) - \mathbf{J}_{d}(\mathbf{z}) \mathbf{f}^r(\mathbf{z}) \Big),
\end{equation}
which can be bounded with the triangle inequality to yield
\begin{equation}\label{eq.DI_error2b}
    \big \lVert \mathbf{J}(\mathbf{u}) \mathbf{f}(\mathbf{u}) - \mathbf{J}_{d}(\mathbf{z}) \mathbf{f}^r(\mathbf{z}) \big \rVert 
    \le 
    \big \lVert \mathbf{J}_{d}(\boldsymbol{\phi}_{e}(\mathbf{u})) \big \rVert
    \cdot
    \big \lVert \mathbf{J}_{e}(\mathbf{u}) \mathbf{f}(\mathbf{u}) 
    - \mathbf{f}^r(\mathbf{z}) \big \rVert
    +
    \big \lVert \mathbf{J}_{d}(\boldsymbol{\phi}_{e}(\mathbf{u})) - \mathbf{J}_{d}(\mathbf{z}) \big \rVert 
    \cdot
    \big \lVert \mathbf{f}^r(\mathbf{z}) \big \rVert
\end{equation}
On the other hand, adding and subtracting $\mathbf{f}(\mathbf{u})$ on the right-hand side of Eq.\eqref{eq.DI_error1} leads to
\begin{equation}\label{eq.DI_error3a}
    \mathbf{J}(\mathbf{u}) \mathbf{f}(\mathbf{u}) - \mathbf{J}_{d}(\mathbf{z}) \mathbf{f}^r(\mathbf{z})
    = \Big( \mathbf{J}(\mathbf{u}) \mathbf{f}(\mathbf{u}) 
    - \mathbf{f}(\mathbf{u}) \Big)
    + \Big( \mathbf{f}(\mathbf{u})  - \mathbf{J}_{d}(\mathbf{z}) \mathbf{f}^r(\mathbf{z}) \Big),
\end{equation}
which is similarly bounded as
\begin{equation}\label{eq.DI_error3b}
    \big \lVert \mathbf{J}(\mathbf{u}) \mathbf{f}(\mathbf{u}) - \mathbf{J}_{d}(\mathbf{z}) \mathbf{f}^r(\mathbf{z}) \big \rVert 
    \le 
    \big \lVert \big( \mathbf{I} - \mathbf{J}(\mathbf{u}) \big) \mathbf{f}(\mathbf{u}) \big \rVert
    +
    \big \lVert \mathbf{f}(\mathbf{u})  - \mathbf{J}_{d}(\mathbf{z}) \mathbf{f}^r(\mathbf{z}) \big \rVert.
\end{equation}
An $\alpha$-linear combination of Eqs. \eqref{eq.DI_error2b} and \eqref{eq.DI_error3b} yields
\begin{equation}\label{eq.DI_error4}
\begin{aligned}
    \big \lVert \mathbf{J}(\mathbf{u}) \mathbf{f}(\mathbf{u}) - \mathbf{J}_{d}(\mathbf{z}) \mathbf{f}^r(\mathbf{z}) \big \rVert 
    & \le 
    \alpha \big \lVert \mathbf{J}_{d}(\boldsymbol{\phi}_{e}(\mathbf{u})) \big \rVert
    \cdot
    \big \lVert \mathbf{J}_{e}(\mathbf{u}) \mathbf{f}(\mathbf{u}) 
    - \mathbf{f}^r(\mathbf{z}) \big \rVert
    +
    \alpha \big \lVert \mathbf{J}_{d}(\boldsymbol{\phi}_{e}(\mathbf{u})) - \mathbf{J}_{d}(\mathbf{z}) \big \rVert 
    \cdot
    \big \lVert \mathbf{f}^r(\mathbf{z}) \big \rVert \\
    & +
    (1-\alpha) \big \lVert \big( \mathbf{I} - \mathbf{J}(\mathbf{u}) \big) \mathbf{f}(\mathbf{u}) \big \rVert
    +
    (1-\alpha) \big \lVert \mathbf{f}(\mathbf{u})  - \mathbf{J}_{d}(\mathbf{z}) \mathbf{f}^r(\mathbf{z}) \big \rVert. 
\end{aligned}
\end{equation}
Supposing that $\mathbf{J}_{d}$ is Lipschitz continuous and bounded, and $\mathbf{f}^r$ is bounded, integration gives
\begin{equation}\label{eq.DI_error_bound}
    \lVert \mathbf{e}_{\text{DI}}(t) \rVert
    \lesssim \int_{t_0}^t \big \lVert \mathbf{J}_{e}(\mathbf{u}) \mathbf{f}(\mathbf{u}) 
    - \mathbf{f}^r(\mathbf{z}) \big \rVert ds 
    + \int_{t_0}^t \big \lVert \boldsymbol{\phi}_{e}(\mathbf{u}) - \mathbf{z} \big \rVert ds
    + \int_{t_0}^t \big \lVert \big( \mathbf{I} - \mathbf{J}(\mathbf{u}) \big) \mathbf{f}(\mathbf{u}) \big \rVert ds
    + \int_{t_0}^t  \big \lVert \mathbf{f}(\mathbf{u})  - \mathbf{J}_{d}(\mathbf{z}) \mathbf{f}^r(\mathbf{z}) \big \rVert ds.
\end{equation} 
Combining the bounds for $\mathbf{e}_{\text{AE}}$ in Eq. \eqref{eq.AE_error_bound} and $\mathbf{e}_{\text{DI}}$ in Eq. \eqref{eq.DI_error_bound}, we obtain the total error estimate
\begin{equation}
    \lVert \mathbf{e}(t) \rVert \lesssim \epsilon_{int}(t;t_0) + \epsilon_{rec}(t;t_0) + \epsilon_{Jac}(t;t_0) + \epsilon_{mod}(t;t_0),
\end{equation}
with
\begin{equation}\label{eq.tlasdi_error_term}
\begin{aligned}
    & \epsilon_{int}(t;t_0) 
     = \int_{t_0}^t \big \lVert \boldsymbol{\phi}_{e}(\mathbf{u}) - \mathbf{z} \big \rVert ds \\
    & \epsilon_{rec}(t;t_0) 
    = \lVert \mathbf{e}_{\text{AE}}(t_0) \rVert
    + \lVert \mathbf{e}_{\text{AE}}(t) \rVert \\
    & \epsilon_{Jac}(t;t_0) 
    = \int_{t_0}^t \big \lVert \big( \mathbf{I} - \mathbf{J}(\mathbf{u}) \big) \dot{\mathbf{u}} \big \rVert ds \\
    & \epsilon_{mod}(t;t_0) 
    = \int_{t_0}^t \big \lVert \mathbf{J}_{e}(\mathbf{u}) \dot{\mathbf{u}} 
    - \mathbf{f}^r(\mathbf{z}) \big \rVert
    + \big \lVert \dot{\mathbf{u}}  - \mathbf{J}_{d}(\mathbf{z}) \mathbf{f}^r(\mathbf{z}) \big \rVert ds.
\end{aligned}
\end{equation}

These quantities correspond directly to continuous counterparts to the loss terms used in training, as defined in Section \ref{sec:loss}.
For instance, applying the trapezoidal approximation over a time interval $[t_n, t_{n+1}]$ gives
\begin{equation}\label{eq.error_int}
\begin{split}
    \epsilon_{int}(t_{n+1};t_n)
    &= \int_{t_n}^{t_{n+1}} \big \lVert \boldsymbol{\phi}_{e}(\mathbf{u}) - \mathbf{z} \big \rVert ds
    \approx \frac{\Delta t}{2} \big \lVert \boldsymbol{\phi}_{e}(\mathbf{u}(t_{n+1})) - \mathbf{z}(t_{n+1}) \big \rVert \\
    & \approx \frac{\Delta t}{2} \Big \lVert \boldsymbol{\phi}_{e}(\mathbf{u}(t_{n+1})) 
    - \boldsymbol{\phi}_{e}(\mathbf{u}(t_{n})) 
    - \int_{t_n}^{t_{n+1}} \mathbf{f}^r(\mathbf{z}(s)) ds \Big \rVert,
\end{split}
\end{equation}
which recovers the structure of the integration loss $\mathcal{L}_{int}$ in Eq. \eqref{eq.loss_ini}.
Similar arguments apply to the remaining terms.
Overall, this error estimate establishes a theoretical upper bound on the tLaSDI-pGFINN approximation error in terms of the training loss components in Eq. \eqref{eq:totalloss}, thereby providing a theoretical foundation for the training strategy.

\subsection{Geometry Conditioned Autoenccoder}
In this study, the Geometry-Conditioned Autoencoder (GC-AE) is proposed to guide the reconstruction process from the latent space to the physical space by incorporating geometry information during decoding process. Some examples in earlier studies including, the DeepONet model \cite{Lu2020} that separates the encoding solution field and the spatial evaluation information through branch and trunk networks, the Denoising Diffusion Model \cite{Graves2024} that embeds geometry conditioning information using Multi-Layer Perceptron (MLP) layers, and the Spatially-Adaptive Denormalization (SPADE) \cite{Park2019}, where semantic spatial information is injected into the decoder through spatially adaptive normalization layers to guide image generation while preserving spatial structure.
To improve the autoencoder's ability to reconstruct the geometry-sensitive damage field, in this study, the geometry information is injected into the decoder architecture using the Feature-Wise Linear Modulation (FiLM) method, which was originally implemented to enhance the visual reasoning task that is challenging for standard deep learning architectures, where question-based information is injected into the extracted input from the CNNs layers \cite{Perez}. FiLM is trained to learn MLP functions $f$ and $h$ which generate feature-level modulation parameters, scaling values $(\boldsymbol{\gamma}_i)$ and shifting values $(\boldsymbol{\beta}_i)$, as functions of input $\boldsymbol{x}_i$: 
\begin{align}
\gamma_{i,c}(\boldsymbol{x}_i) &= f_c(\boldsymbol{x}_i), \\
\beta_{i,c}(\boldsymbol{x}_i) &= h_c(\boldsymbol{x}_i).
\end{align}
where $i$ is the sample index, and $c$ is the feature channel. These modulation parameters $\gamma_{i,c}$ and $\beta_{i,c}$ influence the neural network activations by applying a feature-wise affine transformation to the $c$-th feature map associated with the $i$-th input sample 
\begin{align}
FiLM(\boldsymbol{x}_i)=\gamma_{i,c}(\boldsymbol{x}_i)\circ \boldsymbol{F}_{i,c}(\boldsymbol{x}_i)+\beta_{i,c}(\boldsymbol{x}_i), \end{align} 
where $\boldsymbol{F}_{i,c}$ denotes the activation of the $c$-th feature map corresponding to the $i$-th input sample before the FiLM modulation is applied. This method is shown to perform better than attention based-conditioning \cite{Sun2025} and considered efficient because the same scaling and shifting values, once trained, are injected into every element in the same channel. In addition, FiLM is independent of the input size; hence, increasing the dimension does not significantly increase the computational cost. This is exemplified by the work of \cite{GENG2026131478}, where incorporating FiLM into the diffusion network increased the computational cost by only 3\%. 
The FiLM modulation adaptively scales and shifts the feature activations through the parameters $\gamma_{i,c}$ and $\beta_{i,c}$. Larger $\gamma_{i,c}$ values amplify the corresponding feature activations, enabling the neural network to emphasize more relevant features, while smaller values suppress less informative features during training.  
 
\begin{center}
\refstepcounter{figure}
\includegraphics[width=0.95\textwidth]{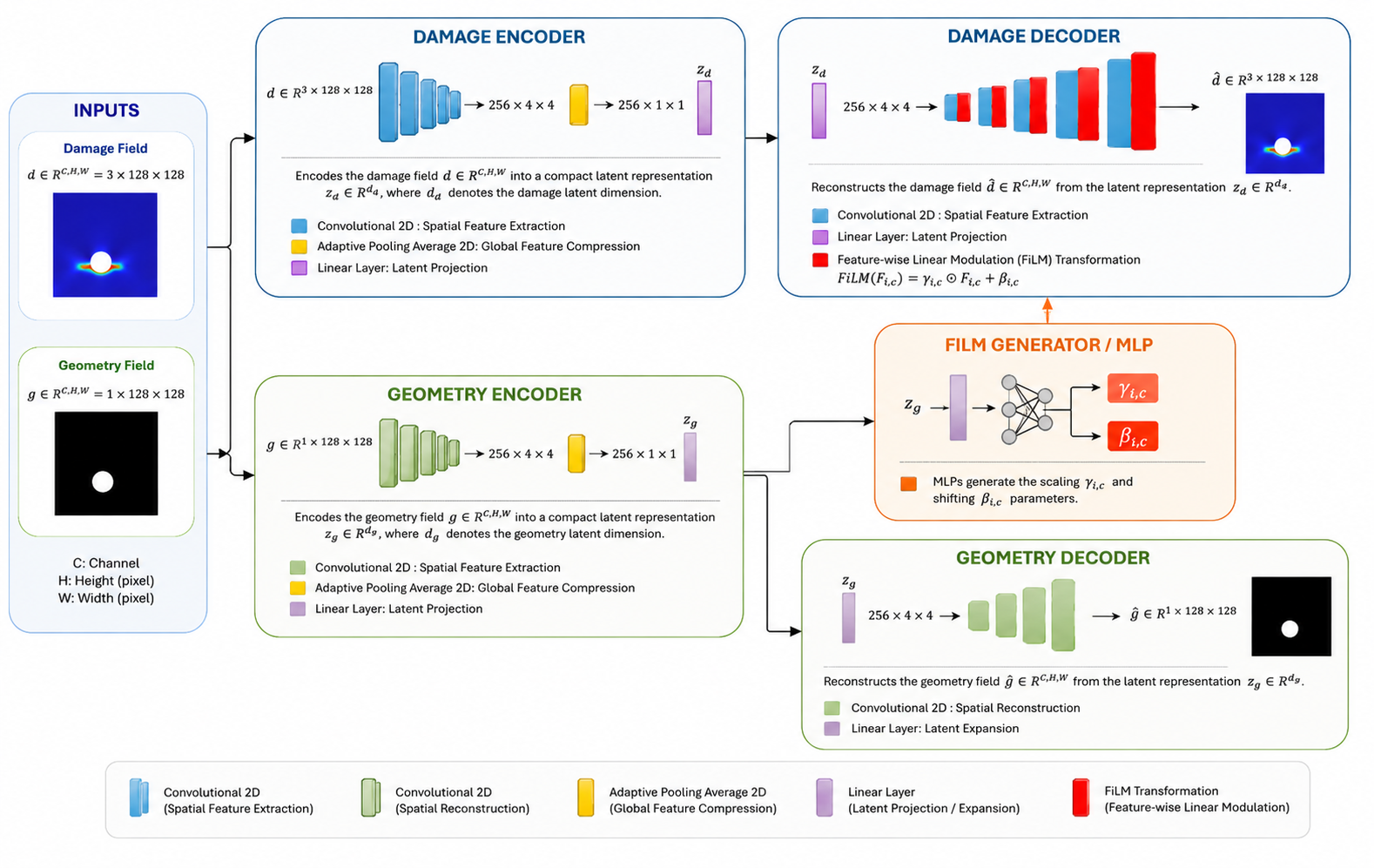}
\\[0.0em]
\textbf{Figure~\thefigure:} Architecture of the Geometry-Conditioned Autoencoder.
\label{fig:Figure7}
\end{center}

In a separate application, \cite{Brockschmidt2020} demonstrated how FiLM is used in GNNs to dynamically modulate node features using information from neighboring source nodes to improve the representation of the target node. This study implements FiLM into the Convolutional Autoencoder, as shown in Figure~\ref{fig:Figure7}. The notation $\mathbb{R}^{C\times H\times W}$ represents the dimensions of input tensor, where $C$ denotes the number of feature channels, $H$ denotes the height of the spatial domain, and $W$ denotes the width of the spatial domain. For example, the geometry input $\mathbb{R}^{1\times128\times128}$ corresponds to a single-channel geometry image with spatial resolution of $128\times128$ pixel points, while the damage input $\mathbb{R}^{3\times128\times128}$ corresponds to a three-channel damage field representation. The linear layers shown in Figure~\ref{fig:Figure7} denote the fully connected NN layers that map extracted feature representations into compact latent vectors. For example, the $\mathbb{R}^{256}\rightarrow\mathbf{z}_g$ represents a linear transformation from a 256-dimensional geometry feature vector into the geometry latent vector, similarly $\mathbb{R}^{256}\rightarrow\mathbf{z}_d$ maps the extracted damage features into the damage latent vector. The ``Reshape'' operation converts a one-dimensional latent vector into a multi-dimensional tensor representation that can be processed by the convolutional decoder layers. The architecture contains two autoencoders that process different inputs, the damage and the geometry, of the surrogate models. Both encoders employ two-dimensional (2D)  five-layer CNNs to compress the spatial dimensions and extract features from the inputs. Subsequently, an adaptive average pooling layer is employed to compress the spatial feature maps (geometry and damage) into compact 1x1 representations for all channels. Average pooling is a down sampling operation commonly used in CNNs to reduce the spatial dimensions of feature maps while reserving important global information. Although the spatial dimensions are reduced, the extracted feature channels retain the global structural and geometric information learned by the encoder. This compact representation enables the latent vector to summarize the overall characteristics of the input field rather than storing explicit pixel-by-pixel information. Then, the damage decoder is integrated with the FiLM layers, where each decoder layer is injected with two modulation parameters to perform the feature-wise modulation. The GC-Autoencoder is trained together with the tLaSDI-pGFINN using the geometric reconstruction loss defined as follows

\begin{equation}
\mathcal{L}_{geom}
=
\lambda_{geom} \sum_{\boldsymbol{\mu}}
\left\|
\mathbf{g}
-
\boldsymbol{\phi}_{d}^g
\!\left(
\boldsymbol{\phi}_{e}^g(\mathbf{g})
\right)
\right\|^2.
\label{eq:loss_rec_geom}
\end{equation}
Here, $\mathbf{g}$, $\boldsymbol{\phi}_{e}^{g}$, and $\boldsymbol{\phi}_{d}^{g}$ denote the geometry field, the geometry encoder, and the geometry decoder, respectively.

Figure~\ref{fig:Figure8} demonstrates the autoencoder reconstructions with and without GC. It is clear that the GC-AE more accurately reconstructs the inclusion geometry and provides improved fracture representation near the inclusion interfaces. In contrast, the model without geometry conditioning produces blurrier reconstructions due to its inability to accurately capture localized damage fields near the material interfaces. In addition, the horizontal crack propagation patterns are more clearly reconstructed by the GC-AE, indicating that the FiLM-generated modulation parameters $(\gamma_i,\ \beta_i)$ effectively guide the decoder to amplify geometry-sensitive fracture features during training.

\begin{center}
\refstepcounter{figure}
\includegraphics[width=0.9\textwidth]{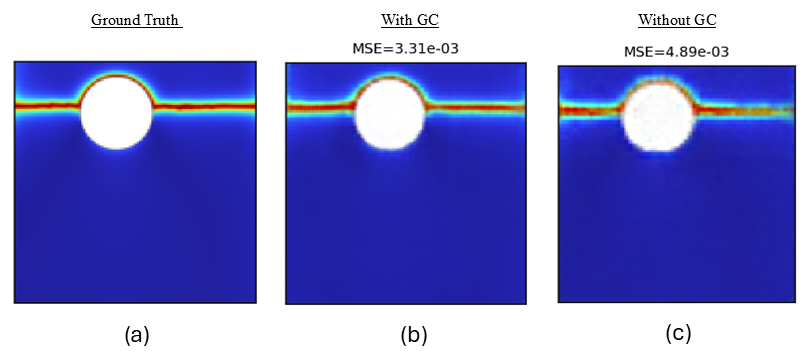}
\\[0.0em]
\textbf{Figure~\thefigure:} Comparison of reconstruction of the damage field: (a) ground truth, (b) prediction with geometry conditioning, (c) prediction without geometry conditioning.
\label{fig:Figure8}
\end{center}

\section{Data Generation and Numerical Setup}
\subsection {Data Generation using Thermodynamically Consistent Phase-Field Model}
For training the proposed thermodynamics-based neural network for fracture prediction, it is important that the training data be generated from simulation models that follow thermodynamic principles. One such example is the phase-field fracture model, where cracks in the material are driven by the elastic energy stored in the body, the surface energy required to propagate the cracks, and the external work. The potential energy is expressed as follows
\vspace{-0.5em}
\begin{equation}
\Pi(\mathbf{u},d)
=
\int_{\Omega}\boldsymbol{\psi}_e(\mathbf{u},d)\,d\Omega
+
\int_{\Omega}\gamma(d)\,g_c\,d\Omega
-
\int_{\Omega}\mathbf{u}\cdot\mathbf{b}\,d\Omega
-
\int_{\Gamma_h}\mathbf{u}\cdot\mathbf{h}\,d\Gamma.
\label{eq:18}
\end{equation}
Here, $\boldsymbol{\psi}_e$ is the elastic energy density, $g_c$ is the critical energy-release rate, $\gamma$ is the fracture surface density function, $d$ is the field parameter, $\mathbf{b}$ and $\mathbf{h}$ are the body forces in domain $\Omega$ and surface traction tensor on traction boundary $\Gamma_h$, respectively. In Phase-field a regularized fracture surface energy is given as:
\vspace{-0.5em}
\begin{equation}
\gamma
=
\frac{1}{2l}d^2
+
\frac{l}{2}|\boldsymbol{\nabla}d|^2,
\label{eq:19}
\end{equation}
where $l$ is the regularization length scale. Cracks generally occur due to the tensile energy; hence, the elastic energy density is decomposed into two parts, tensile and compressive:
\vspace{-0.5em}
\begin{equation}
\psi_e
=
\left((1-d)^2(1-k)+k\right)\psi^{+}
+
\psi^{-},
\label{eq:20}
\end{equation}
where $\psi^{+}$ and $\psi^{-}$ are the tensile and compressive strain energies with respect to the principal strain decomposition, respectively, $(1-d)^2$ is the degradation factor of the $\psi^{+}$ elastic energy, and $k \ll 1$ is a small positive number introduced to maintain the well-posedness of the system.

In the standard phase-field model, $\mathbf{u}$ and $d$ are obtained by minimization of $\Pi(\mathbf{u},d)$ in Eq.~(\ref{eq:18}). In \cite{Baek2024} and \cite{Baek2022}, a modified procedure is considered, where $d$ is related to $\mathbf{u}$ via $\delta\Pi(\mathbf{u},d)=0$ with consideration $\gamma=\frac{1}{2l}d^2$. This leads to
\vspace{-0.5em}
\begin{equation}
d(\boldsymbol{x},t)
=
\frac{2H(\boldsymbol{x},t)}
     {2H(\boldsymbol{x},t)+g_c/l}.
\label{eq:21}
\end{equation}
\vspace{-1.0em}

For irreversible fracture process, the following $H(\boldsymbol{x},t)$ is employed for the phase-field $d$ in Eq.~(\ref{eq:21})

\vspace{-0.5em}
\begin{equation}
H(\boldsymbol{x},t)
=
\max\left\{\psi^{+}(\boldsymbol{x},\tau)\right\},
\quad
\tau \in [0,t]
\label{eq:22}
\end{equation}
\vspace{-0.5em}

The solution $\mathbf{u}$ and $d$ are solved by (\ref{eq:18}) and (\ref{eq:21}) in a staggered manner.

\subsection {Baseline Model}
In this study, Autoencoder-LSTM (AE-LSTM) is employed as an baseline model to compare the prediction performance of the thermodynamics-based model. LSTM is a type of recurrent neural network (RNN) that learns complex input-output relationships by capturing the temporal dependencies from the sequential data. The typical RNN architecture is shown in 
Figure~\ref{fig:RNN}.
\begin{center}
\refstepcounter{figure}
\includegraphics[width=0.3\textwidth]{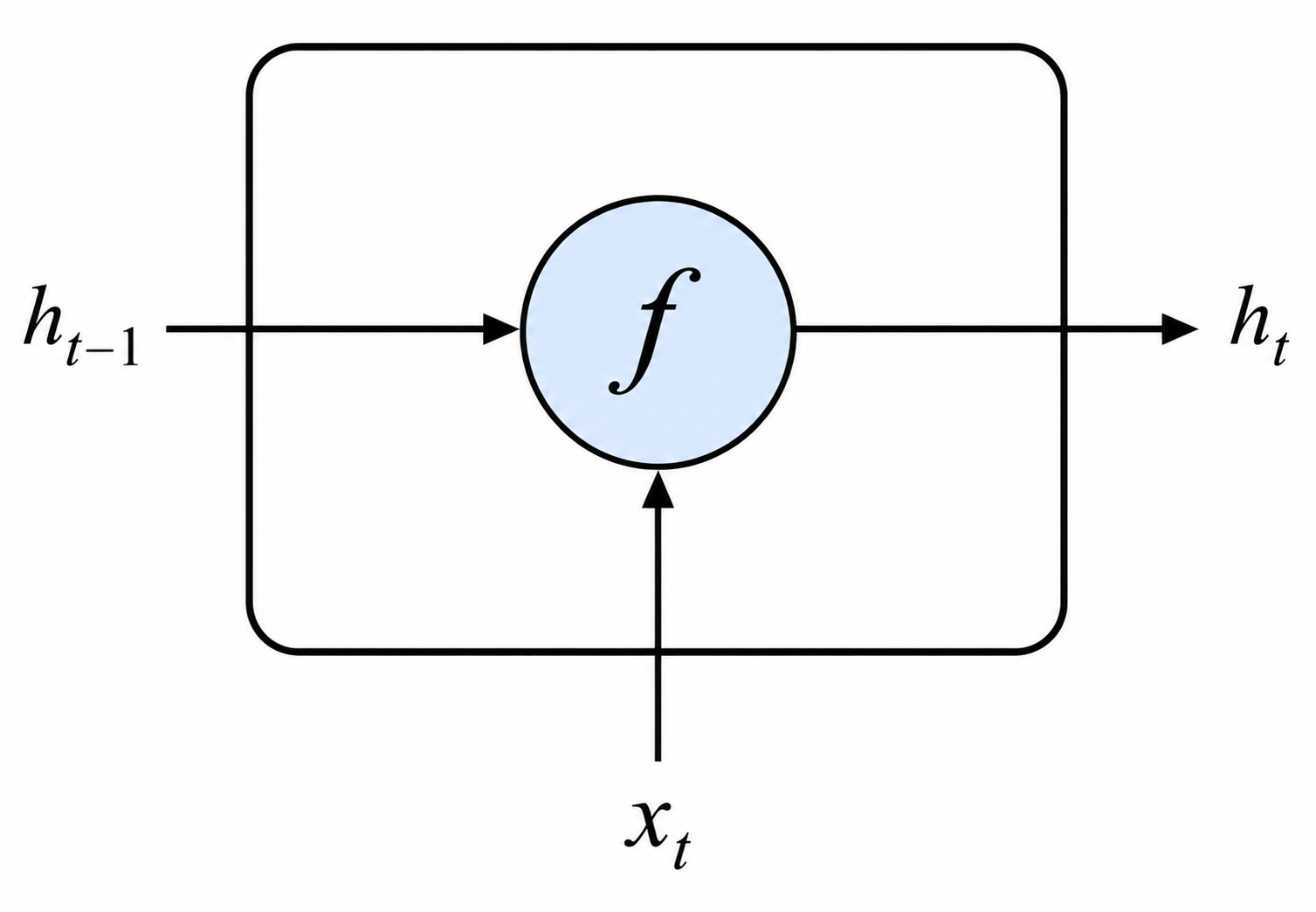}
\\[0.0em]
\textbf{Figure~\thefigure:} A simple architecture of a Recurrent Neural Network.
\label{fig:RNN}
\end{center}

The sequential RNN learning process is given as follows.
\vspace{-0.5em}
\begin{equation}
\mathbf{h}_t=f_{\mathrm{act}}(\mathbf{h}_{t-1},\mathbf{x}_t).
\end{equation}

Here, $\mathbf{x}_t$ and $\mathbf{h}_t$ are the input and the hidden state, storing information at time $t$, with $f_{\mathrm{act}}$ denotes the activation function. For standard RNNs, the tanh activation function is employed, as expressed in Eq.(4), where $\mathbf{w}_h$ and $\mathbf{w}_x$ are the trainable weight parameters associated with the hidden state $\mathbf{h}_{t-1}$ and input $\mathbf{x}_t$, respectively, and $\mathbf{b}_h$ is the bias.
\begin{equation}
\mathbf{h}_t
=\tanh\left(
\mathbf{w}_h\mathbf{h}_{t-1}
+
\mathbf{w}_x\mathbf{x}_t
+
\mathbf{b}_h
\right).
\end{equation}

However, as the data sequence lengthens, standard RNNs fail to learn long-range dependencies. This issue arises from two difficulties associated with backpropagation through time: vanishing gradients, which prevent meaningful gradient updates for earlier inputs, and exploding gradients, where gradients grow excessively large, leading to unstable training \cite{Yu2019}.

\begin{center}
\refstepcounter{figure}
\includegraphics[width=0.55\textwidth]{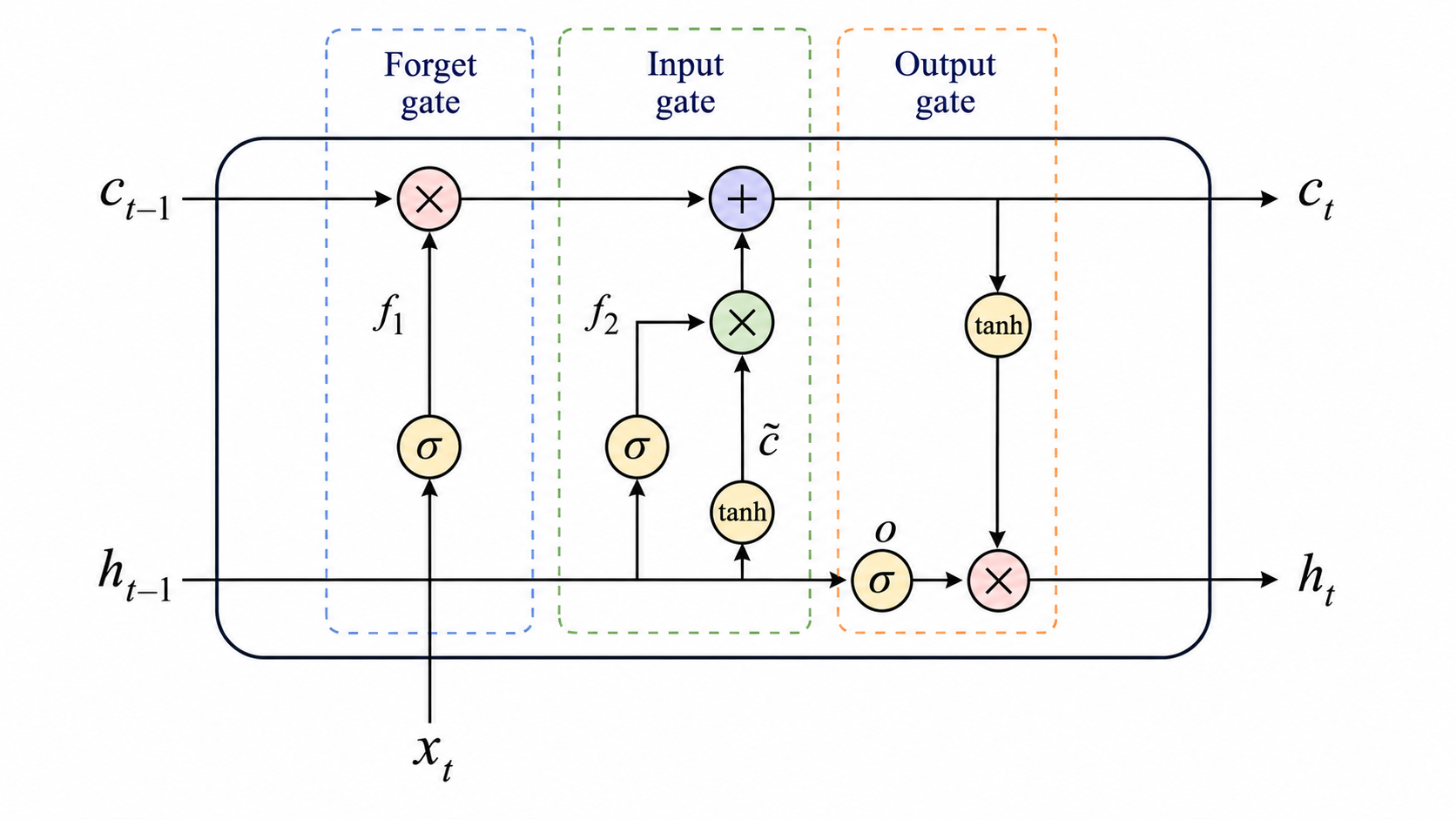}
\\[0.0em]
\textbf{Figure~\thefigure:} Architecture of Long-Short Term Memory.
\label{fig:LSTM}
\end{center}

LSTM, as shown in Figure~\ref{fig:LSTM}, addresses the issue of long-term dependencies by incorporating gates into the cell. One of these gates is the forget gate $f_1$ which decides which information needs to be removed from the cell state with the sigmoid activation function $\sigma$. When the forget value is 1, the information is retained, otherwise the information is removed.
\begin{equation}
f_1=\sigma\left(
\mathbf{w}_{1h}\mathbf{h}_{t-1}
+
\mathbf{w}_{1x}\mathbf{x}_t
+
\mathbf{b}_{f_1}
\right).
\label{eq:input_gate}
\end{equation}

The Input gate determines which new information should be stored in the long-term memory. This gate consists of two parts: the gate controller in Eq.~\eqref{eq:gate_cont} which determines how much information should be included in the memory, and the candidate memory in Eq.~\eqref{eq:candidate_m} , which determines what information enters the memory.
\vspace{-0.5em}
\begin{equation}
f_2=\sigma\left(
\mathbf{w}_{2h}\mathbf{h}_{t-1}
+
\mathbf{w}_{2x}\mathbf{x}_{t}
+
\mathbf{b}_{f_2}
\right).
\label{eq:gate_cont}
\end{equation}
\vspace{-2.0em}
\begin{equation}
\tilde{c}=\tanh\left(
\mathbf{w}_{ch}\mathbf{h}_{t-1}
+
\mathbf{w}_{\tilde{c}x}\mathbf{x}_{t}
+
\mathbf{b}_{\tilde{c}}
\right).
\label{eq:candidate_m}
\end{equation}

The third gate is the output gate, which decides which internal memory is passed to the next sequence.
\vspace{-0.50em}
\begin{equation}
o=\sigma\left(
\mathbf{w}_{oh}\mathbf{h}_{t-1}
+
\mathbf{w}_{ox}\mathbf{x}_t
+
\mathbf{b}_{o}
\right).
\label{eq:output_gate}
\end{equation}

From Eq.~\eqref{eq:input_gate} to Eq.~\eqref{eq:output_gate}, a cell state can be computed as follows,
\vspace{-0.5em}
\begin{equation}
c_t=f_1\cdot c_{t-1}+f_2\cdot\tilde{c}.
\end{equation}

A new hidden state is obtained by multiplying the output gate by the hyperbolic tangent of the cell state,
\vspace{-0.5em}
\begin{equation}
\mathbf{h}_t=o\cdot\tanh(c_t).
\end{equation}

In this work, the AE-LSTM model is trained using the reconstruction loss defined in Eq. \eqref{eq:loss_rec}. Figure~\ref{fig:AELSTM} illustrates the inference workflow of AE-LSTM. The high-dimensional physical state at time step $n$, denoted by $\mathbf{u}_n$, is encoded into a low-dimensional latent representation $\mathbf{z}_n$ using the convolutional encoder. The LSTM takes $\mathbf{z}_n$ as input and predicts the latent state at the subsequent time step, $\mathbf{z}_{n+1}$. The convolutional decoder then maps $\mathbf{z}_{n+1}$ back into the physical space, yielding the predicted fracture field $\tilde{\mathbf{u}}_{n+1}$. The dashed loop in Figure~\ref{fig:AELSTM} represents the autoregressive prediction process, in which the predicted fracture field $\tilde{\mathbf{u}}_{n+1}$ is re-encoded by the convolutional encoder to generate the latent representation for the next prediction step.

\begin{center}
\refstepcounter{figure}
\includegraphics[width=0.95\textwidth]{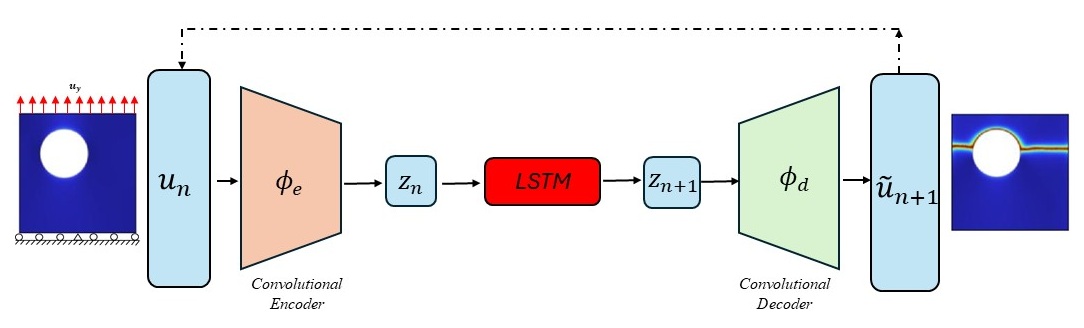}
\\[0.0em]
\textbf{Figure~\thefigure:} Schematic of Autoencoder-LSTM (AE-LSTM) in inference mode.
\label{fig:AELSTM}
\end{center}

\section{Numerical Results}
This chapter discusses the training and testing results of the machine learning models. All models are run on the PyTorch platform \citep{paszke2019pytorch} with the Adam optimizer \citep{kingma2014adam}. Model training is conducted on the COSMOS high-performance computing (HPC) platform at the San Diego Supercomputer Center, utilizing AMD Instinct MI300A APUs.
To quantitatively evaluate the prediction accuracy of the proposed framework, two error metrics are employed throughout this section: the mean squared error (MSE) and the mean relative $\ell_2$ (RelL2) error. The MSE measures the average squared difference between the predicted and reference fracture fields over all spatial pixels and time steps for each testing sample, and is subsequently averaged over all testing samples. It is defined as
\begin{equation}
\mathrm{MSE}
=
\frac{1}{N_s}
\sum_{k=1}^{N_s}
\left(
\frac{1}{N_p}
\left|\left|
\tilde{\mathbf{U}}^{(k)}-\mathbf{U}^{(k)}
\right|\right|_F^2
\right),
\label{eq:mse}
\end{equation}
where $\mathbf{U}^{(k)}=\Big[\mathbf{u}_{0}^{(k)}, \mathbf{u}_{1}^{(k)}, .., \mathbf{u}_{N_t}^{(k)}\Big]$ and $\tilde{\mathbf{U}}^{(k)}=\Big[\tilde{\mathbf{u}}_{0}^{(k)}, \tilde{\mathbf{u}}_{1}^{(k)}, .., \tilde{\mathbf{u}}_{N_t}^{(k)}\Big]$ denote the reference and predicted fracture fields of the $k$-th testing sample, respectively, with $N_t$ as the total number of time steps; $N_p$ is the total number of values (all spatial pixels over all time steps) in a single testing sample, and $N_s$ is the total number of testing samples.

The mean relative $\ell_2$ error measures the prediction error relative to the magnitude of the reference solution for each testing sample and is subsequently averaged over all testing samples. It is defined as
\begin{equation}
\mathrm{RelL2}=
\frac{1}{N_s}
\sum_{k=1}^{N_s}
\frac{
\left|\left|
\tilde{\mathbf{U}}^{(k)}-\mathbf{U}^{(k)}
\right|\right|_F
}{
\left|\left|
\mathbf{U}^{(k)}
\right|\right|_F
},
\label{eq:rell2}
\end{equation}
where $\left|\left|\cdot\right|\right|_F$ denotes the Frobenius norm.

\subsection{Single Cavity Fracture Evolution under Varying External Tractions}

The first numerical example considers a phase-field fracture dataset generated from simulations of a square domain containing a cavity  subjected to tensile loading on its outer boundaries \cite{Hakimzadeh2025}, as shown in Figure~\ref{fig:cavity_domain}. The non-dimensionalized material properties are defined as $\mu = 1$, $G_c = 1.72 \times 10^{-5}$ and the material is assumed to exhibit no Poisson effect. Eight short cracks are pre-defined on the internal traction-free cavity surface. The phase-field fracture model with a neo-Hookean hyperelastic constitutive law was used to simulate crack propagations under varying external tensile tractions as reported in \cite{Hakimzadeh2025} and the results are used for model training. The dataset is divided into training, validation, and testing sets, consisting of 800, 50, and 50 data samples, respectively. Each sample consists of 45 time steps describing the fracture evolution on an 80×80-pixel spatial grid. Prescribed displacements on the outer boundary $\Gamma$ are imposed as

\vspace{-0.5em}
\begin{equation}
\mathbf{u}=\mathbf{\bar{F}}\mathbf{x},
\end{equation}
or
\vspace{-0.5em}
\begin{equation}
\begin{pmatrix}
u_1 \\
u_2
\end{pmatrix}
=
\begin{pmatrix}
F_{11} & F_{12} \\
F_{21} & F_{22}
\end{pmatrix}
\begin{pmatrix}
x_1 \\
x_2
\end{pmatrix},
\end{equation}
where

\vspace{-1.0em}
\begin{equation}
\begin{aligned}
\mathbf{\bar{F}}(\mathbf{x},t) &= \alpha(\mathbf{x},t)\,\mathbf{I},\\
\end{aligned}
\end{equation}
with $\mathbf{x}\in\Gamma$, $0.0 \le t \le 1.0$, and  $\alpha(\mathbf{x},t)$ is a Gaussian Random Field (GRF). This results in spatially varying affine boundary conditions.

\begin{center}
\refstepcounter{figure}
\includegraphics[width=0.35\textwidth]{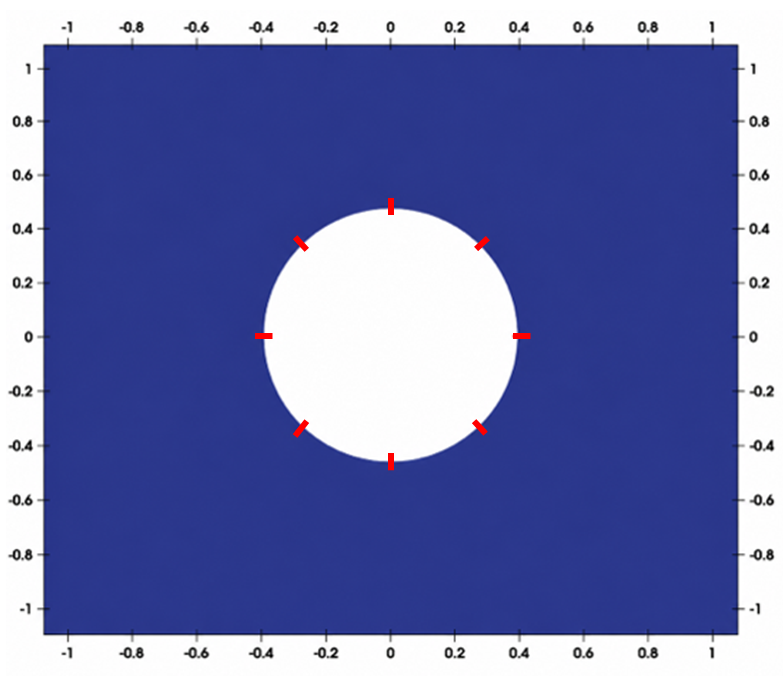}
\\[0.0em]
\textbf{Figure~\thefigure:} 2D Domain with fixed inclusion and eight pre-cracks (red tick marks denote initial cracks).
\label{fig:cavity_domain}
\end{center}

The autoencoder maps the high-dimensional damage field to a five-dimensional latent representation through encoder layers with channel dimensions of 16, 64, 128, and 256 architecture. ReLU activation functions are applied in the encoder except the final layer, while the decoder adopts a symmetric architecture. The detailed network architecture is illustrated in Figure~\ref{fig:Figure7}. A geometry autoencoder is unnecessary in this example because the geometry is simple and invariant across all samples.

For the tLaSDI-pGFINN model, the pGFINN consists of 10 fully connected hidden layers with 20 neurons per layer. The latent dynamics are advanced using the Forward Euler integration scheme. The model is trained using the combined objective consisting of the tLaSDI-pGFINN loss in Eq.~\eqref{eq:totalloss} and the geometric reconstruction loss in Eq.~\eqref{eq:loss_rec_geom}, with regularization parameters \(\lambda_{rec}=10^{-1}\), \(\lambda_{jac}=10^{-9}\), and \(\lambda_{mod}=10^{-7}\). The initial learning rate is set to \(10^{-5}\) and decayed by a factor of 0.99 every 1000 epochs. 

The baseline model, AE-LSTM, employs the same autoencoder architecture. The model is initialized with a learning rate of \(10^{-3}\) and trained using the reconstruction loss. A ReduceLROnPlateau scheduler is employed to adapt the learning rate during training, where the learning rate is reduced by a factor of 0.5 if the validation loss does not improve for twenty five consecutive epochs, with a minimum learning rate of \(10^{-5}\).
The training and validation histories of tLaSDI-pGFINNs with Geometry Conditioner (GC) are shown Figure~\ref{fig:Figure10}. The loss curve shows no significant oscillations during training, indicating a stable learning process. The curve converges after 30,000 epochs, suggesting that additional iterations yield minimal improvement. Furthermore, no evidence of overfitting is observed, and the final training and validation losses remain within the same order of magnitude.

\begin{center}
\refstepcounter{figure}
\includegraphics[width=0.45\textwidth]{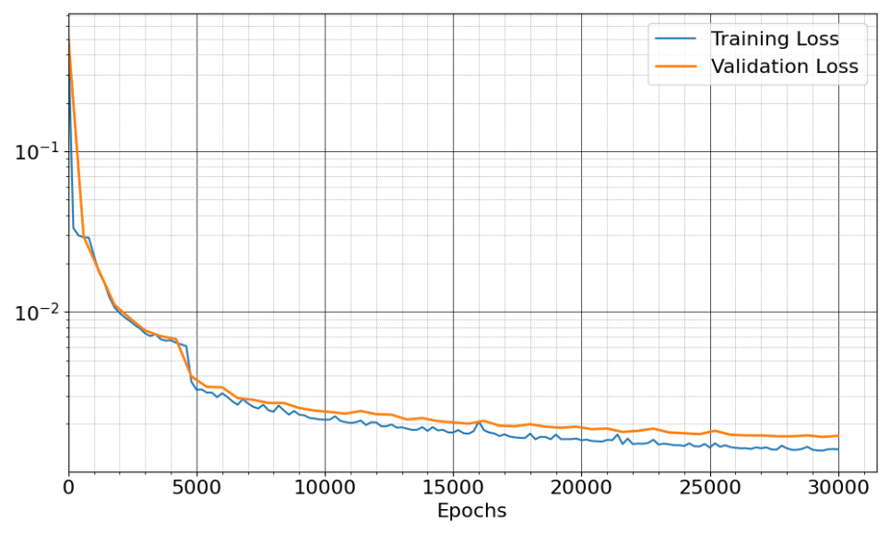}
\\[0.0em]
\textbf{Figure~\thefigure:} Training and validation loss histories of tLaSDI-pGFINN.
\label{fig:Figure10}
\end{center}

After training, the model's generalization ability to predict fracture patterns is evaluated on unseen samples (testing). An autoregressive approach is adopted, where the prediction result of the current time step is used as input to predict the subsequent time step. This way, the model’s ability to predict the full crack evolution can be assessed. The average of MSE and RelL2 defined in Eq.~\eqref{eq:mse} and Eq.~\eqref{eq:rell2}  error across all samples in the testing set is summarized in Table~\ref{tab:test_metrics}. Figure~\ref{fig:Figure11} shows the bar plots of MSE from all samples in the testing set. It can be observed that 70\% of the samples are predicted more accurately by the tLaSDI-pGFINNs compared to the AE-LSTM, indicating that the thermodynamics-based model exhibits better generalization capability on unseen testing samples.

\begin{table}[!htbp]
\centering
\refstepcounter{table}
\label{tab:test_metrics}

{\bfseries Table \thetable. Prediction performance on the Testing Dataset}

\vspace{0.3em}

\begin{tabular}{lcc}
\hline
& \multicolumn{2}{c}{Model} \\
\cline{2-3}
Metric & AE-LSTM & tLaSDI-pGFINN \\

\hline
MSE & $1.877\times10^{-3}$ & $1.411\times10^{-3}$ \\
Average L2 Error & 12.47\% & 10.68\% \\
\hline
\end{tabular}

\end{table}

\begin{center}
\refstepcounter{figure}
\includegraphics[width=0.85\textwidth]{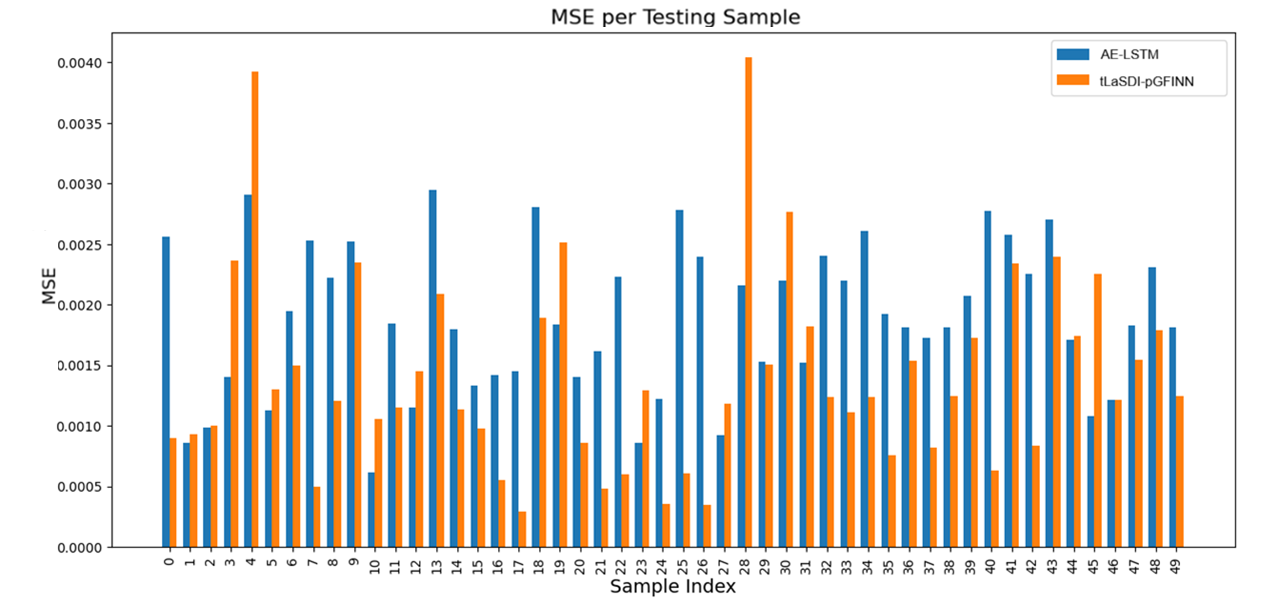}
\makebox[\textwidth][c]{\textbf{Figure~\thefigure:} Comparison of MSE for each testing sample between AE-LSTM and tLaSDI-pGFINN.}
\label{fig:Figure11}
\end{center}

Figure~\ref{fig:Figure12}-Figure~\ref{fig:Figure16} depict the prediction results of five samples in the testing set. Across these samples, tLaSDI-pGFINNs demonstrates higher accuracy in predicting crack propagation over time. In contrast, the AE-LSTM produces inaccurate crack evolution. This accuracy in crack evolution prediction is due to the fact that pGFINN is trained to learn the state variables of the damage at every time step during training by inferring the energy and entropy conditions. On the other hand, the LSTM relies heavily on the cell state, which stores information from the previous step, and this process of determining which information to retain is purely driven by nonlinear gating mechanisms using sigmoid and hyperbolic tangent activation functions without physics guidance. Although about 30\% of the samples are still predicted more accurately by AE-LSTM, this model requires substantially longer sequences and larger input data to match the accuracy of tLaSDI-pGFINN when forecasting fractures in unseen samples under an autoregressive setting where only the initial input is provided.


\begin{center}
\refstepcounter{figure}
\includegraphics[width=\textwidth]{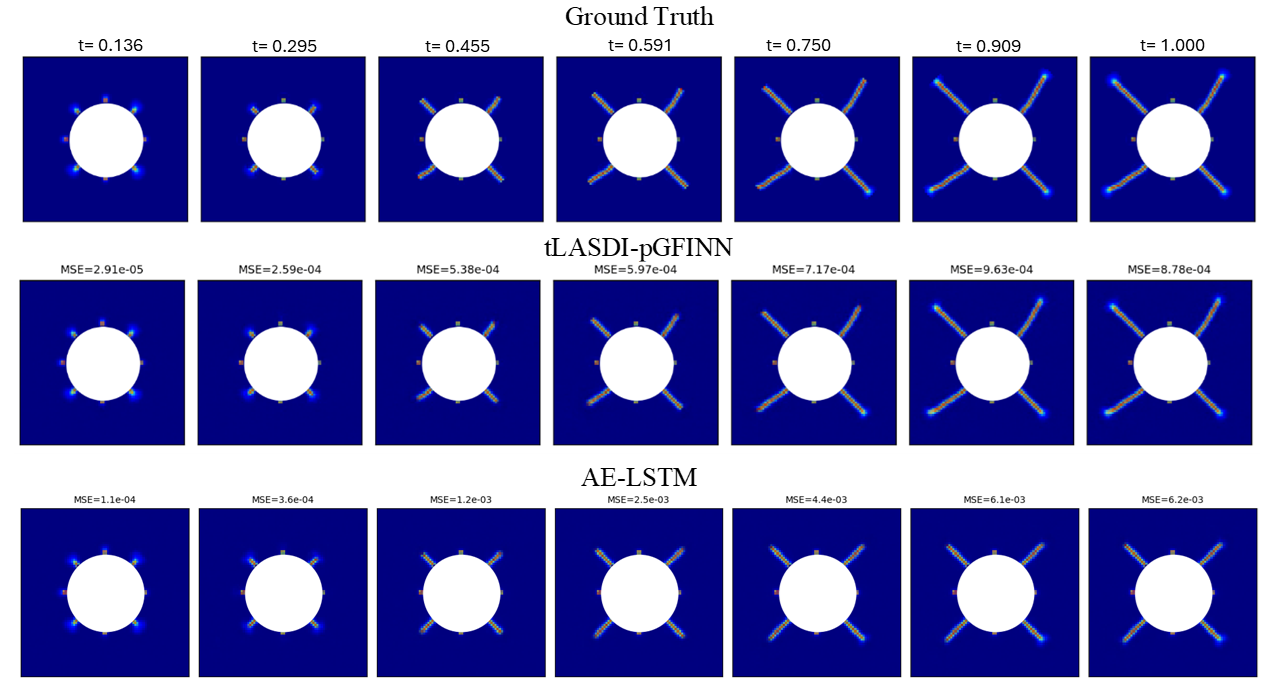}
\\[-1.5em]
\textbf{Figure~\thefigure:} tLaSDI-pGFINN vs AE-LSTM prediction for testing sample 7.
\label{fig:Figure12}
\end{center}

\begin{center}
\refstepcounter{figure}
\includegraphics[width=\textwidth]{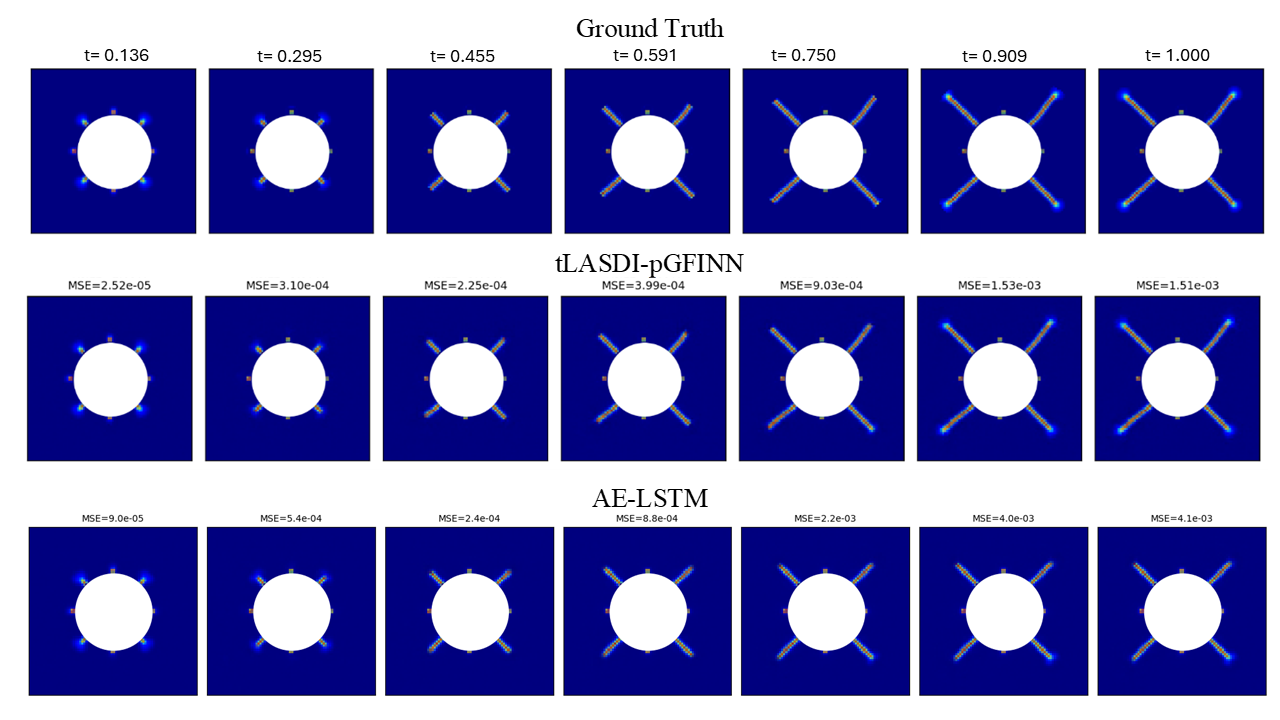}
\\[-1.0em]
\textbf{Figure~\thefigure:} tLaSDI-pGFINN vs AE-LSTM prediction for testing sample 16.
\label{fig:Figure13}
\end{center}

\begin{center}
\refstepcounter{figure}
\includegraphics[width=\textwidth]{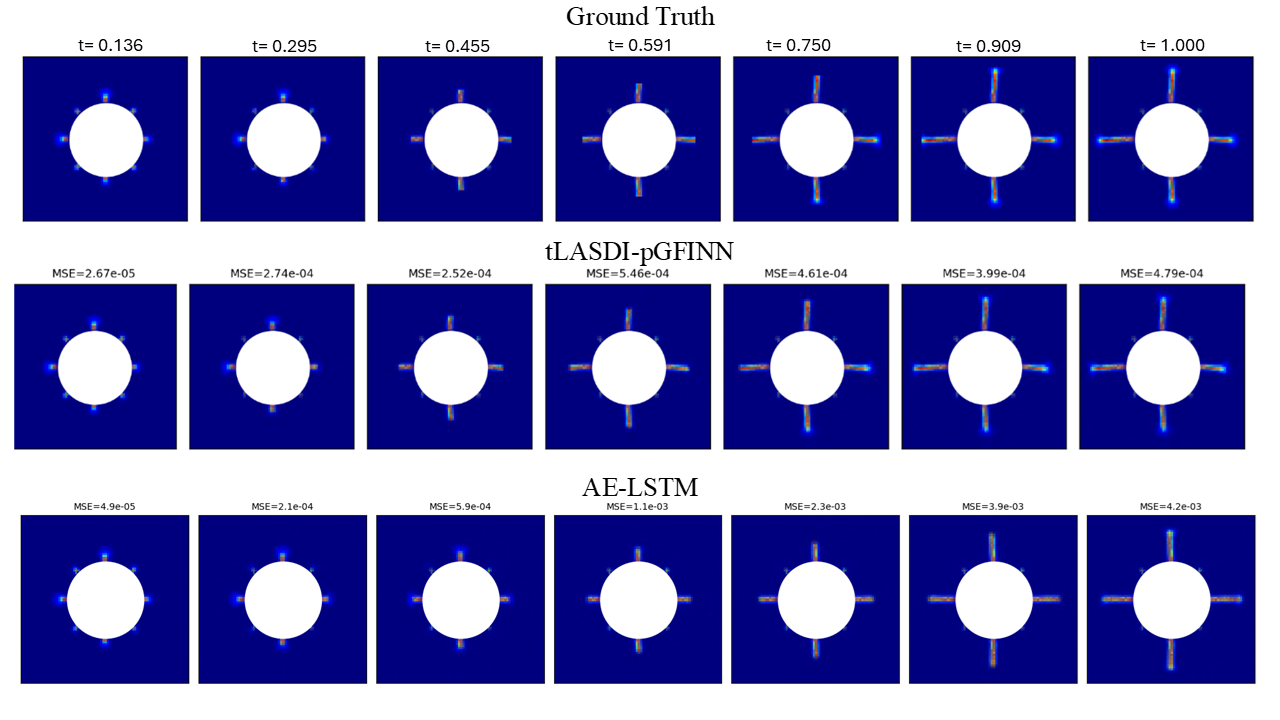}
\\[-1.5em]
\textbf{Figure~\thefigure:} tLaSDI-pGFINN vs AE-LSTM prediction for testing sample 17.
\label{fig:Figure14}
\end{center}

\begin{center}
\refstepcounter{figure}
\includegraphics[width=\textwidth]{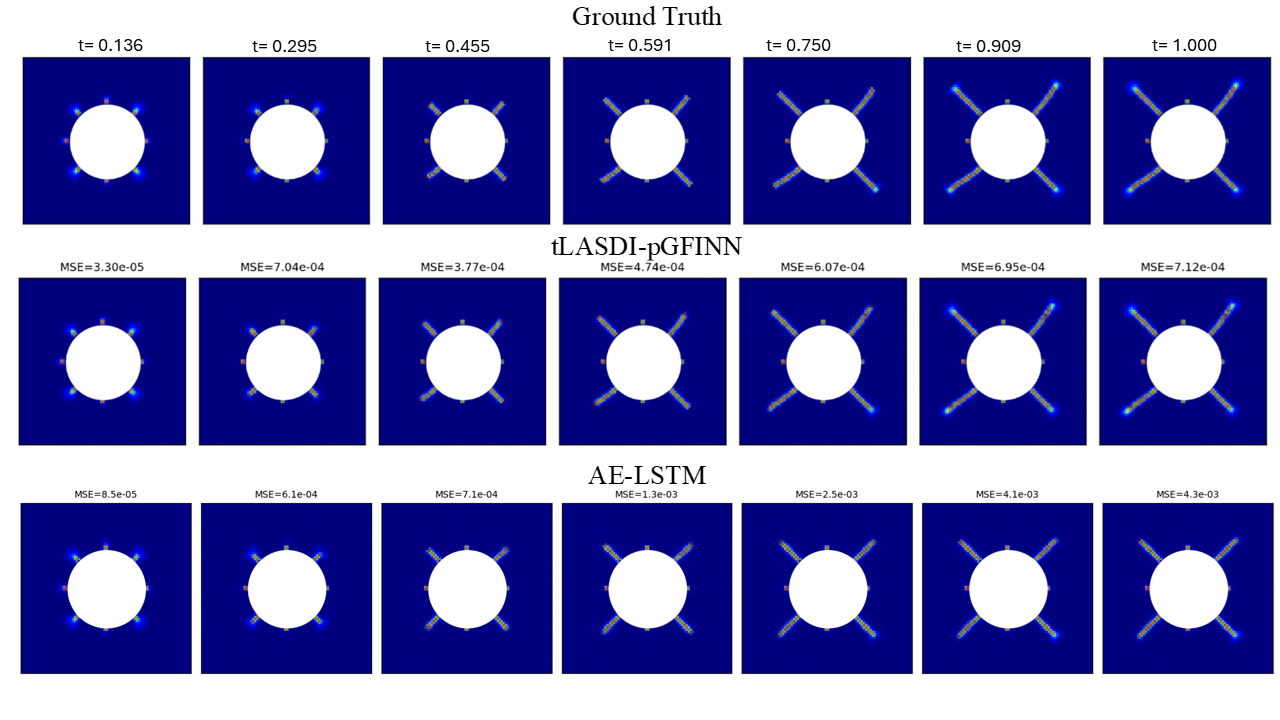}
\\[-1.5em]
\textbf{Figure~\thefigure:} tLaSDI-pGFINN vs AE-LSTM prediction for testing sample 21.
\label{fig:Figure15}
\end{center}
 
\begin{center}
\refstepcounter{figure}
\includegraphics[width=\textwidth]{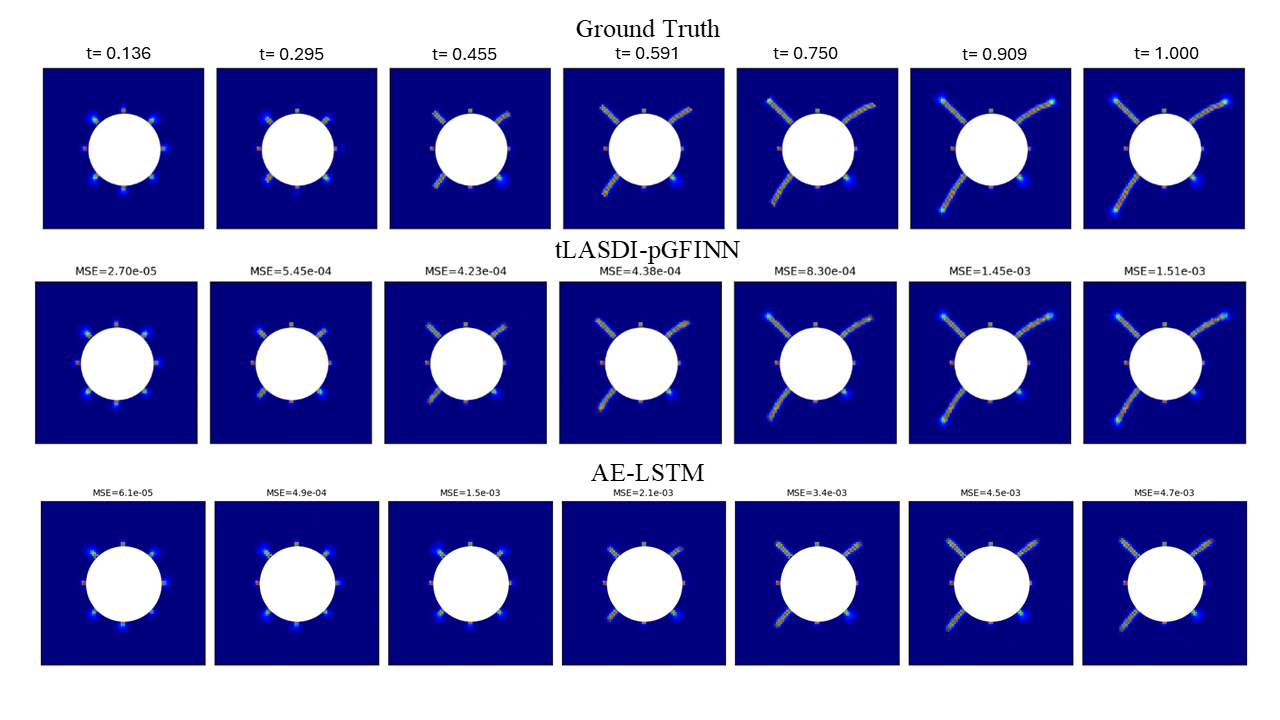}
\\[-1.5em]
\textbf{Figure~\thefigure:} tLaSDI-pGFINN vs AE-LSTM prediction for testing sample 39.
\label{fig:Figure16}
\end{center}
 
Furthermore, to quantify how well the thermodynamic conditions are followed during the prediction, the energy and entropy histories over all time steps of the tLaSDI-pGFINN model are evaluated. Figure~\ref{fig:Figure17} shows the energy and entropy history plots. The energy history plot demonstrates the conservation of energy over time. Throughout the time-evolution, the energy time rate remains on the order of $10^{-7}$ at all time steps, demonstrating the robustness in imposing physics in pGFINN. As the energy required to propagate fracture is thermodynamically linked to the entropy generation, it is essential observing the non-decreasing entropy production, as shown in second row of Figure~\ref{fig:Figure17}. These results demonstrate that tLaSDI-pGFINN contributes significantly to accurate crack-evolution prediction, owing to its ability to encode the thermodynamic conditions that regulate fracture behavior. Table~\ref{tab:timing} summarizes the training and prediction times of both models. AE-LSTM exhibits a shorter training time due to its relatively simple, purely data-driven architecture. In contrast, tLaSDI-pGFINN incurs higher training costs because of the enforcement of thermodynamic constraints and the Jacobian vector product computation in the loss functions. Nevertheless, once trained, tLaSDI-pGFINN achieves higher accuracy, while requiring only approximately 5 seconds per test case for prediction.

\begin{table}[!htbp]
\centering
\refstepcounter{table}
\label{tab:timing}

{\bfseries Table \thetable. Training and Prediction Times of AE-LSTM and tLaSDI-pGFINN}

\vspace{0.3em}

\begin{tabular}{lcc}
\hline
Model & Training & Prediction \\
      & Time & Time \\
\hline
AE-LSTM & $\sim$15 mins & $\sim$3 s \\
tLaSDI-pGFINN & $\sim$6 Hours & $\sim$5 s \\
\hline
\end{tabular}

\end{table}

\begin{center}
\refstepcounter{figure}
\includegraphics[width=0.9\textwidth]{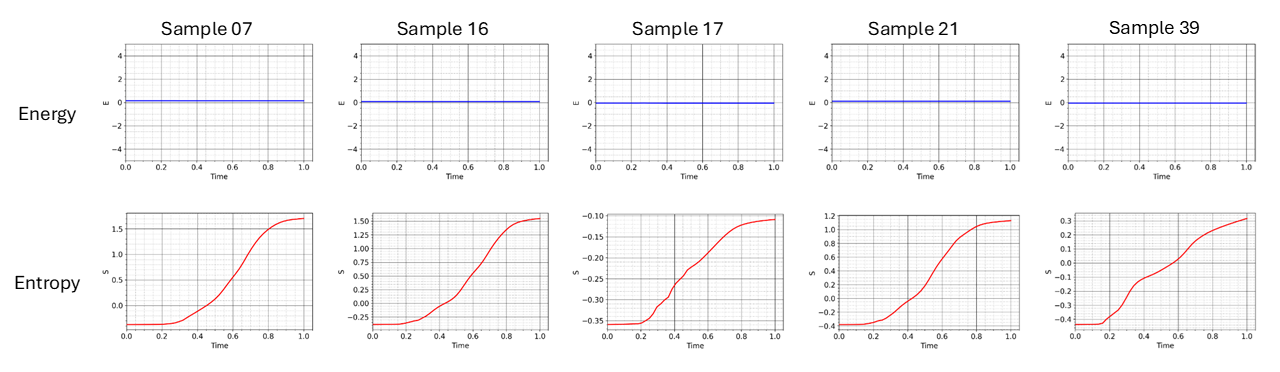}
\vspace{0.5em}
\makebox[\textwidth][c]{\textbf{Figure~\thefigure:} Energy and entropy histories predicted by pGFINN.}
\label{fig:Figure17}
\end{center}

\subsection{Single-Inclusion Fracture with Geometry Variation}

The second numerical example evaluates the capability of the proposed tLaSDI-pGFINN framework to predict crack evolution in heterogeneous materials with increased parametric variability. A phase-field-generated dataset consisting of 400 samples is employed, where each sample contains a single inclusion embedded within the domain with varying diameters and spatial locations in both the horizontal and vertical directions. Each sample consists of 48 time steps describing the fracture evolution on an 120×120-pixel spatial grid. The young’s modulus, Poisson’s ratio and fracture energy of inclusion are set to 52000 MPa, 0.3, and 0.1 N/m, respecitvely. For the matrix, the young’s modulus, Poisson’s ratio and fracture energy are set to 10400 MPa, 0.3 and 0.1 N/m, respectively. The material is subjected to a prescribed vertical displacement $u_y(t)=0.013t$ mm, $0.0 \leq t \leq 1.0$ on the top surface as shown in Figure~\ref{fig:Figure18}.

\begin{center}
\refstepcounter{figure}
\includegraphics[width=0.25\textwidth]{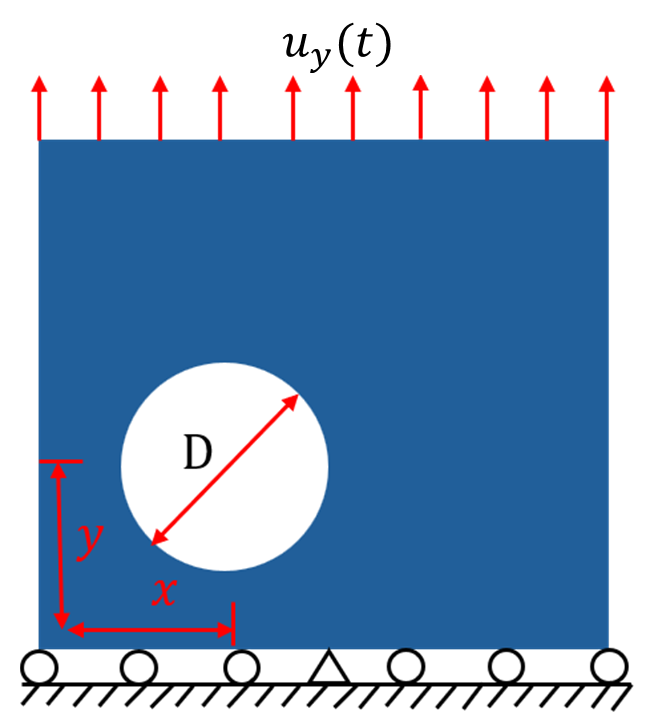}

\textbf{Figure~\thefigure:} Numerical model with vertical displacement prescribed on the top surface of the material domain.
\label{fig:Figure18}
\end{center}

The damage encoder consists of four hidden layers with channel dimensions of 16, 32, 64, and 128 followed by a five-dimensional latent space. The geometry encoder consists of four hidden layers with channel dimensions of 8, 16, 32, and 64, followed by a 32-dimensional latent space. ReLU activation functions are employed throughout the encoders, except the final layer, while the decoders adopt symmetric architectures.

The latent dynamics are modeled using a pGFINN comprising 10 fully connected hidden layers with 40 neurons per layer. The latent state evolution is integrated using the Forward Euler scheme. During training, the loss function defined in Eq.~\eqref{eq:totalloss} is minimized with regularization coefficients \(\lambda_{rec}=10^{-1}\), \(\lambda_{jac}=10^{-10}\), and \(\lambda_{mod}=10^{-9}\). An initial learning rate of \(10^{-5}\) is adopted and reduced by a factor of 0.99 every 1000 epochs. 
For comparison, an AE-LSTM model is constructed using the same encoder-decoder architecture. The model is optimized using only the reconstruction loss and initialized with a learning rate of \(10^{-3}\). A ReduceLROnPlateau scheduler is employed, reducing the learning rate by a factor of 0.5 when the validation loss does not improve for twenty five consecutive epochs, subject to a minimum learning rate of \(10^{-5}\).

The training and validation histories of tLaSDI-pGFINN are shown in Figure~\ref{fig:Figure19}. The plot shows stable training behavior where no significant oscillations appear during the training. No overfitting is observed, as the validation loss does not diverge or increase while the training loss decreases.

\begin{center}
\refstepcounter{figure}
\includegraphics[width=0.45\textwidth]{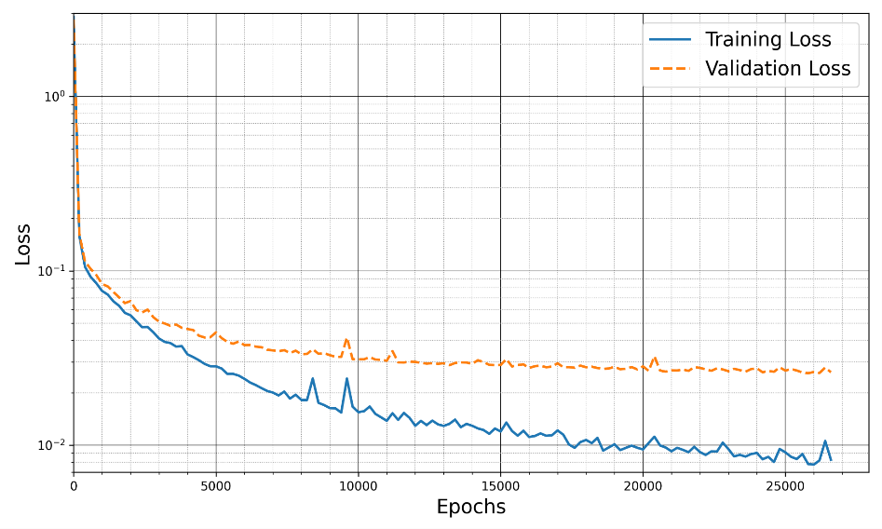}

\vspace{0.5em}

\textbf{Figure~\thefigure:} Training and validation loss histories of tLaSDI-pGFINN.
\label{fig:Figure19}
\end{center}

Table~\ref{tab:test_metrics2} and Figure~\ref{fig:Figure20} summarize the model performance on the testing set. It is evident that the tLaSDI-pGFINN model with Geometry Conditioner outperforms the AE-LSTM in this study.

\begin{table}[!htbp]
\centering
\refstepcounter{table}
\label{tab:test_metrics2}

{\bfseries Table \thetable. Prediction performance on the Testing Dataset}

\vspace{0.3em}

\begin{tabular}{lcc}
\hline
& \multicolumn{2}{c}{Model} \\
\cline{2-3}
Metric & AE-LSTM & tLaSDI-pGFINN \\
& w/ GC & w/GC \\
\hline
MSE & $1.17\times10^{-2}$ & $7.50\times10^{-3}$ \\
Average L2 Error & 18.40\% & 15.67\% \\
\hline
\end{tabular}

\end{table}

\begin{center}
\refstepcounter{figure}
\includegraphics[width=0.9\textwidth]{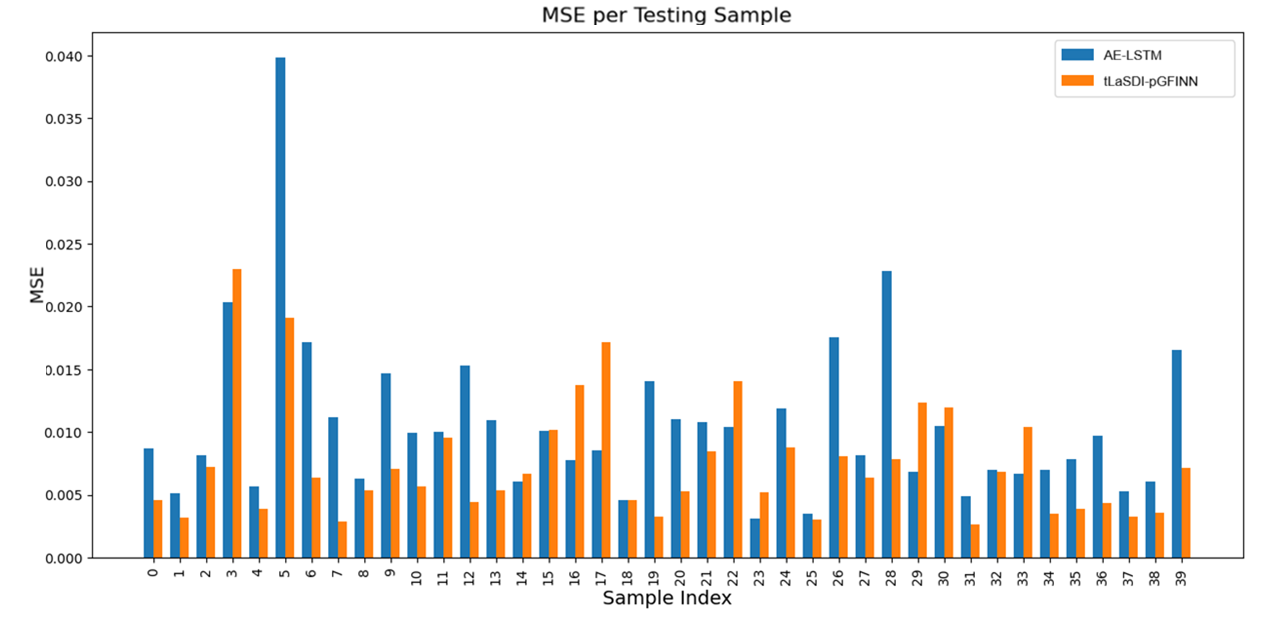}

\vspace{0.5em}

\textbf{Figure~\thefigure:} Comparison of MSE for each testing sample between AE-LSTM and tLaSDI-pGFINN.
\label{fig:Figure20}
\end{center}

To further visualize the model performance, Figure~\ref{fig:Figure21} to Figure~\ref{fig:Figure25} show the progressive crack evolution histories. The results demonstrate that tLaSDI-pGFINN model generates sharper and more accurate crack pattern predictions than those generated by AE-LSTM. The feature-level modulation mechanism in the Geometry-Conditioned Architecture allows the tLaSDI-pGFINN model to learn relevant features within the heterogeneous material subdomain geometries, particularly for the cases where the crack initiation are strongly affected by the heterogeneous subdomain geometries.

As can be seen, incorporating thermodynamic constraints into the tLaSDI-pGFINN model improves its generalization capability. Figure~\ref{fig:Figure21} and Figure~\ref{fig:Figure22} clearly demonstrate cases where the AE-LSTM model fails to accurately predict the fracture evolution across the time steps. In contrast, the tLaSDI-pGFINN model is able to predict the crack evolution much more closely to the ground truth. This behavior is particularly evident in Figure~\ref{fig:Figure21}, where crack initiation occurs at a later time step, and the tLaSDI-pGFINN model still captures the crack initiation accurately.

In Figure~\ref{fig:Figure23}, the inclusion is located very close to the boundary, representing a extrapolation condition. In this case, the AE-LSTM fails to reproduce the fully developed crack pattern at the end of the simulation, while tLaSDI-pGFINN’s prediction remains relatively accurate in predicting the crack evolution throughout the entire time history.

Similarly, in Figure~\ref{fig:Figure24} the tLaSDI-pGFINN can reasonably capture the crack evolution at each time step, while the AE-LSTM reaches the fully developed crack state prematurely at time step $t=0.702$. Although there are cases such as Figure~\ref{fig:Figure25}, where both models can predict the fracture evolution reasonably close to the ground truth, it is important to note that, on average the tLaSDI-pGFINN model demonstrates better consistency and robustness in predicting unseen samples compared to the AE-LSTM, particularly in capturing the state of crack initiation and the overall morphology of the crack evolution. These observations highlight the importance of incorporating physical constraints into the machine learning framework. In particular, the thermodynamic constraints improve the stability and reliability of autoregressive rollout predictions, where the accuracy at each future time step strongly depends on the predictions from previous time steps.
 
\begin{center}
\refstepcounter{figure}
\includegraphics[width=0.99\textwidth]{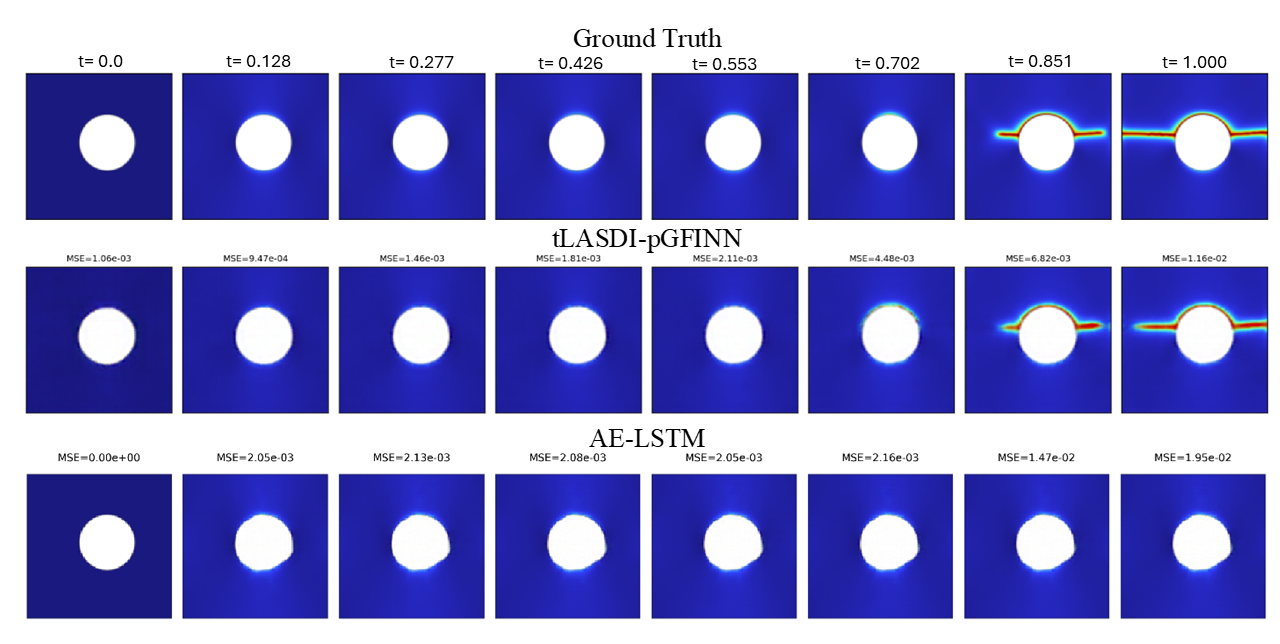}
\\[-1.0em]
\textbf{Figure~\thefigure:} tLaSDI-pGFINN vs AE-LSTM prediction for testing sample 38.
\label{fig:Figure21}
\end{center}

\begin{center}
\refstepcounter{figure}
\includegraphics[width=0.99\textwidth]{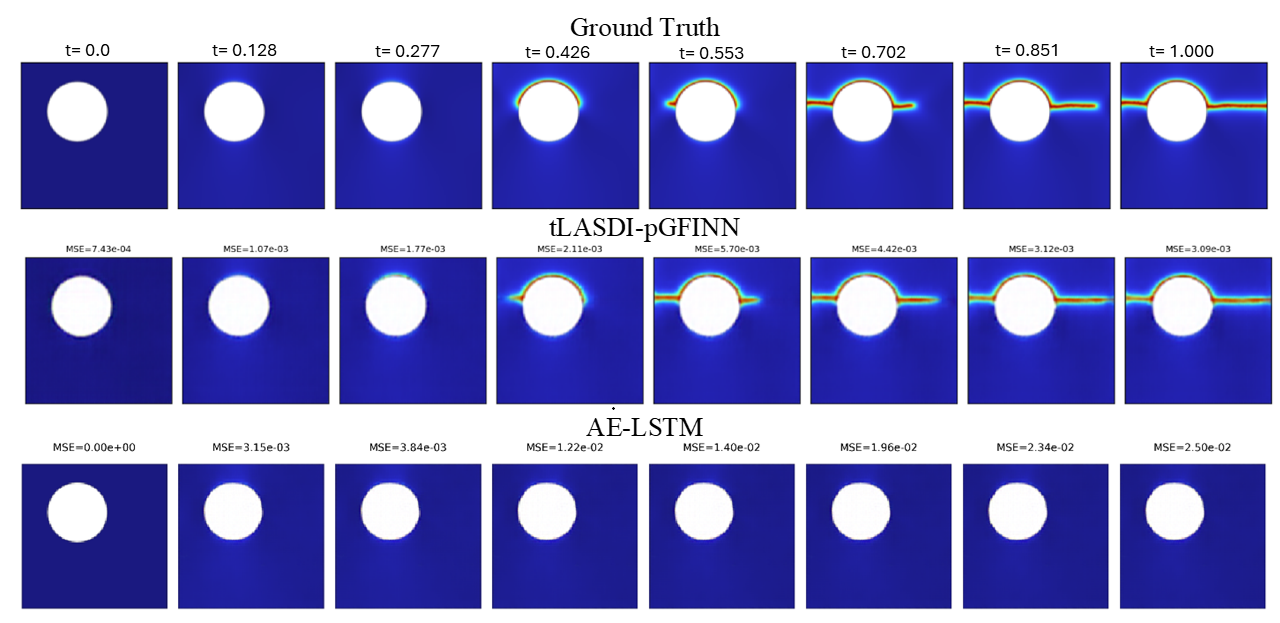}
\\[-0.80em]
\textbf{Figure~\thefigure:} tLaSDI-pGFINN vs AE-LSTM prediction for testing sample 19.
\label{fig:Figure22}
\end{center}

\begin{center}
\refstepcounter{figure}
\includegraphics[width=0.99\textwidth]{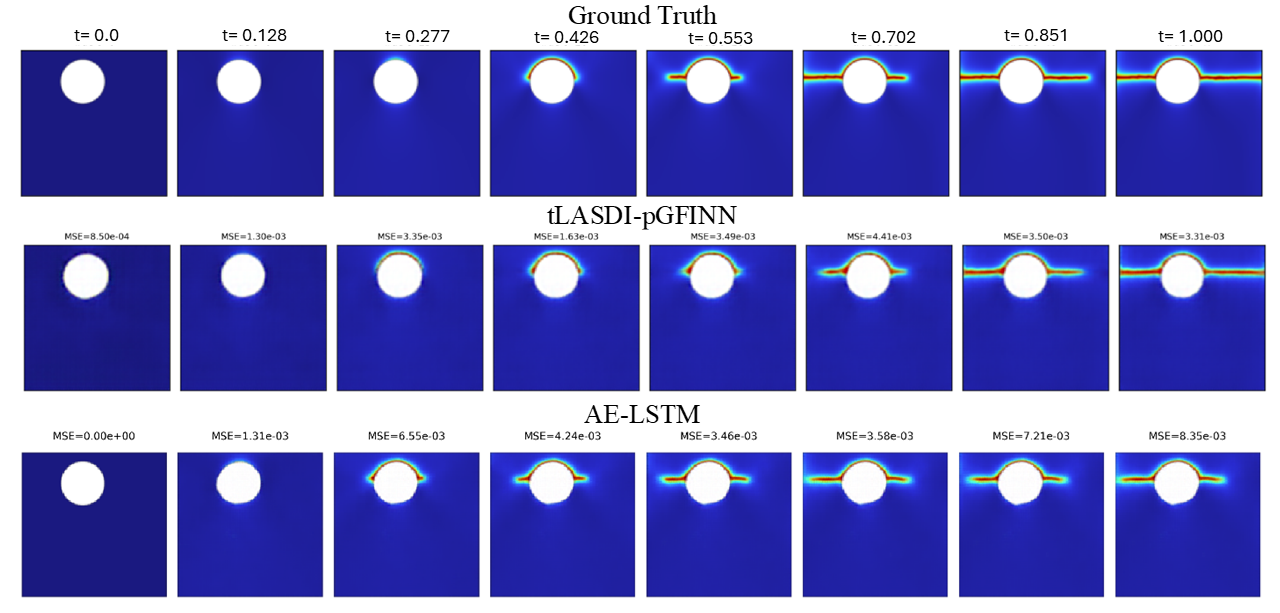}
\\[-1.0em]
\textbf{Figure~\thefigure:} tLaSDI-pGFINN vs AE-LSTM prediction for testing sample 31.
\label{fig:Figure23}
\end{center}

\begin{center}
\refstepcounter{figure}
\includegraphics[width=0.99\textwidth]{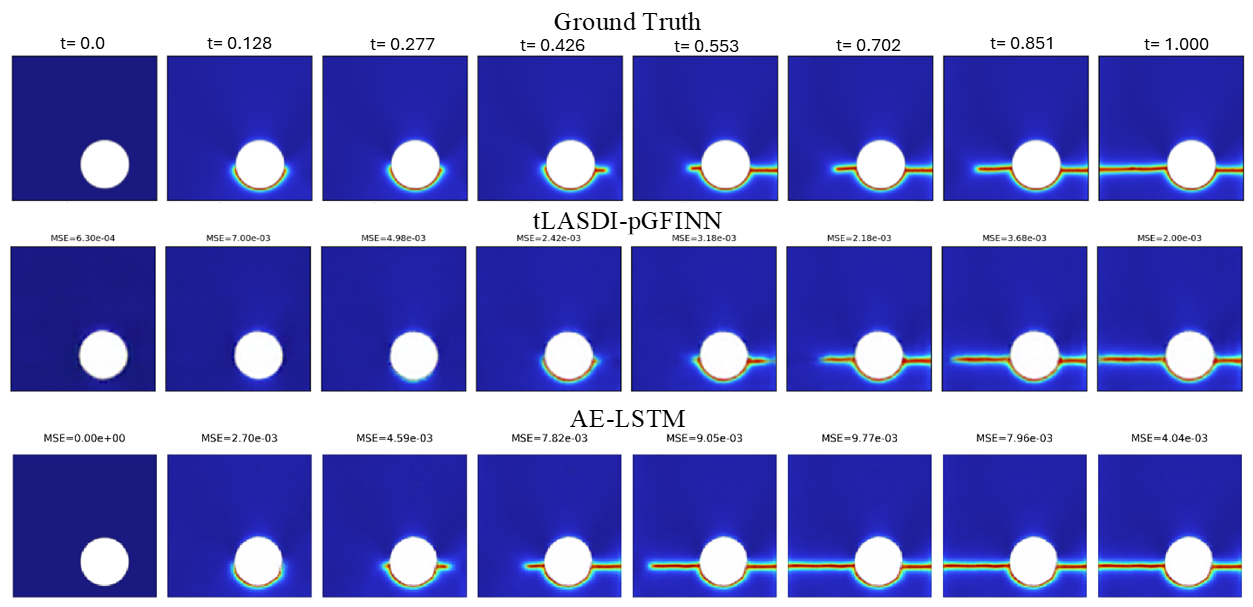}
\\[-0.7em]
\textbf{Figure~\thefigure:} tLaSDI-pGFINN vs AE-LSTM prediction for testing sample 34.
\label{fig:Figure24}
\end{center}

\begin{center}
\refstepcounter{figure}
\includegraphics[width=0.99\textwidth]{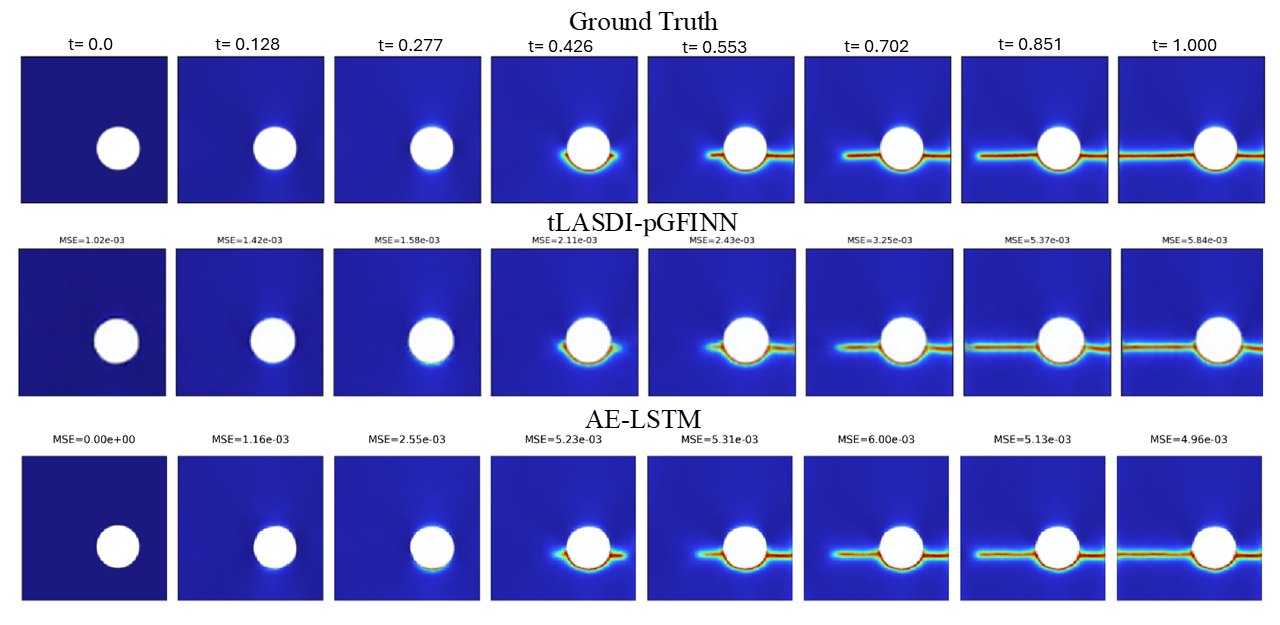}
\\[-1.0em]
\textbf{Figure~\thefigure:} tLaSDI-pGFINN vs AE-LSTM prediction for testing sample 1.
\label{fig:Figure25}
\end{center}

Figure~\ref{fig:Figure26} shows the energy and entropy time histories. These plots demonstrate that the tLaSDI-pGFINN model preserves the thermodynamic constraints even for unseen testing samples. The energy plot remains nearly constant throughout the crack evolution meeting the energy conservation. In addition, the monotonic increase in entropy indicating the irreversible fracture process throughout the loading history. The entropy plots also provide insight into the crack formation behavior over time. For example, sample 38 exhibits a relatively small entropy rate from $t = 0.2$ to $t = 0.6$, indicating that no significant dissipation occurs during this period. This observation is consistent with the crack evolution results, where crack initiation is only observed after time step $t = 0.553$. In contrast, sample 31 shows crack initiation at an earlier time step $t=0.277$, which is supported by the higher entropy rate observed from $t = 0.0$ to $t = 0.4$. These observations further emphasize the importance of thermodynamic constraints in accurately predicting crack evolution over time.

Table~\ref{tab:timing2} summarizes the training and prediction times of both models. The incorporation of the geometry-conditioned autoencoder architecture increases the training time for both models. However, this additional training cost leads to improve prediction accuracy. Furthermore, once the models are trained, tLaSDI-pGFINN achieve higher accuracy, while requiring only approximately 5 seconds per test case. Compared with the full-order simulation that requires approximately 36 min (2160 s), the tLaSDI-pGFINN model achieves a speedup of approximately 432 times.

\begin{center}
\refstepcounter{figure}
\includegraphics[width=0.95\textwidth]{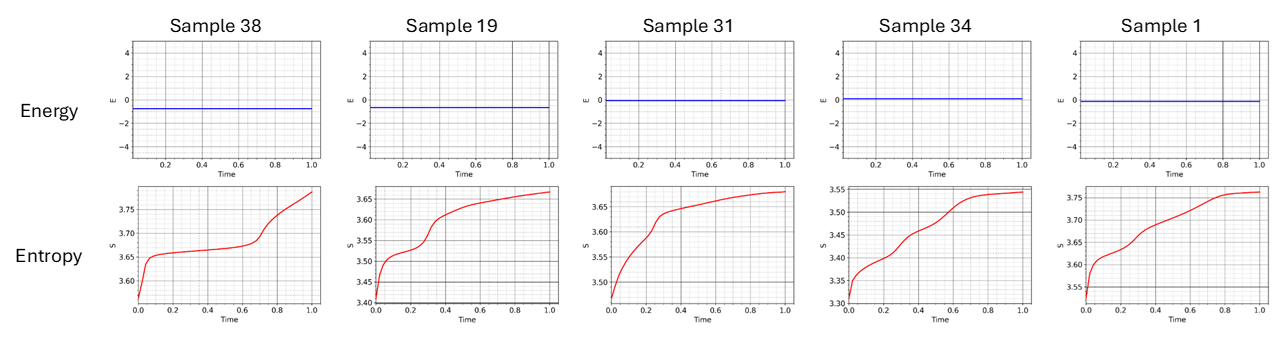}
\\[-0.0em]
\textbf{Figure~\thefigure:} Energy and entropy histories predicted by pGFINN.
\label{fig:Figure26}
\end{center}

\begin{table}[!htbp]
\centering
\caption{Training and Prediction times of the full-order model, AE-LSTM, and tLaSDI-pGFINN.}
\label{tab:timing2}

\begin{tabular}{lccc}
\hline
Model & Training Time & Prediction Time & Speedup \\
\hline
Full-Order Simulation & -- & $\sim$36 min & $--$ \\
AE-LSTM & $\sim$20 min & $\sim$3 s & $\sim720\times$ \\
tLaSDI-pGFINN & $\sim$8 h & $\sim$5 s & $\sim432\times$ \\
\hline
\end{tabular}
\end{table}

\subsection{Multiple-Inclusion Fracture with Varying Crack Initiation Locations}

In this example, crack propagation in a two-dimensional domain containing multiple inclusions is investigated. Figure~\ref{fig:Figure27} illustrates the problem setup according to \cite{hakimzadeh2025data}.A prescribed vertical displacement $u_y(t)=1.2\times10^{-6}\,t~\mathrm{m}$, $0.0 \leq t \leq 1.0$, is applied on the top surface $\Gamma$ of the problem domain. The pre-notch crack is randomly located on the left vertical surface of the specimen. The Young’s modulus, Poisson’s ratio, and fracture energy of the inclusions are set to 3770 GPa, 0.3, and 0.1 N/m, respectively, whereas the corresponding material properties of the matrix are 377 GPa, 0.3, and 0.01 N/m. A total of 500 samples is employed to train the model, where 80\% of the data is used for training, and the remaining 20\% is used for validation and testing. Each sample consists of 30 time steps describing the fracture evolution on an 128×64-pixel spatial grid.
\begin{center}
\refstepcounter{figure}
\includegraphics[width=0.16\textwidth]{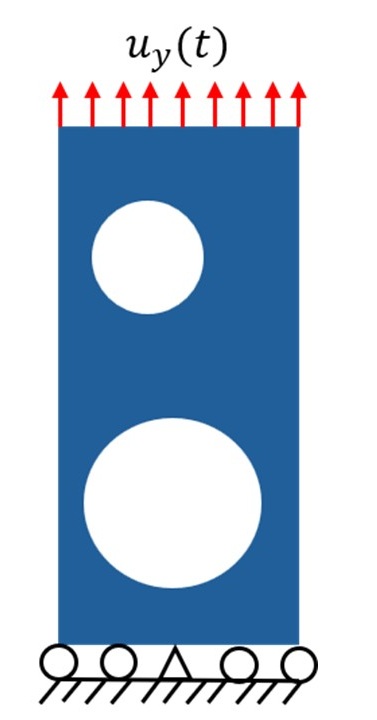}
\\[0.0em]
\textbf{Figure~\thefigure:} Problem setup with multiple inclusions and a random pre-notch crack on the left vertical surface.
\label{fig:Figure27}
\end{center}

The autoencoder maps the high-dimensional damage field to a five-dimensional latent representation through encoder layers with channel dimensions of 16, 32, 64, and 128. The geometry autoencoder consists of four hidden layers with channel dimensions of 8, 16, 32, and 64, followed by a 32-dimensional latent space. ReLU activation functions are applied to all encoder layers except the final layer. The decoder adopts a symmetric architecture to reconstruct the physical field from the latent variables, as illustrated in Figure~\ref{fig:Figure7}.

The pGFINN used to evolve the latent dynamics consists of 10 hidden layers with 20 neurons per layer. Time integration is performed using the Forward Euler method. Model training is carried out using the objective function defined in Eq.~\eqref{eq:totalloss}, with regularization coefficients \(\lambda_{rec}=10^{-1}\), \(\lambda_{jac}=10^{-7}\), and \(\lambda_{mod}=10^{-7}\). The learning rate is initially set to \(10^{-5}\) and gradually reduced by 1\% every 1000 epochs.  

To benchmark the predictive performance, an AE-LSTM model sharing the same autoencoder architecture is considered. The network is trained using the reconstruction loss only. Optimization begins with a learning rate of \(10^{-3}\), and a ReduceLROnPlateau scheduler decreases the learning rate by a factor of 0.5 when no improvement in the validation loss is observed for twenty five consecutive epochs, with a minimum learning rate of \(10^{-5}\).  

The training loss plot of tLaSDI pGFINN is shown Figure~\ref{fig:Figure28}. No overfitting is observed, and the training and validation losses converge within the same order of magnitude, with final loss value of approximately $\sim10^{-2}$.
\begin{center}
\refstepcounter{figure}
\includegraphics[width=0.45\textwidth]{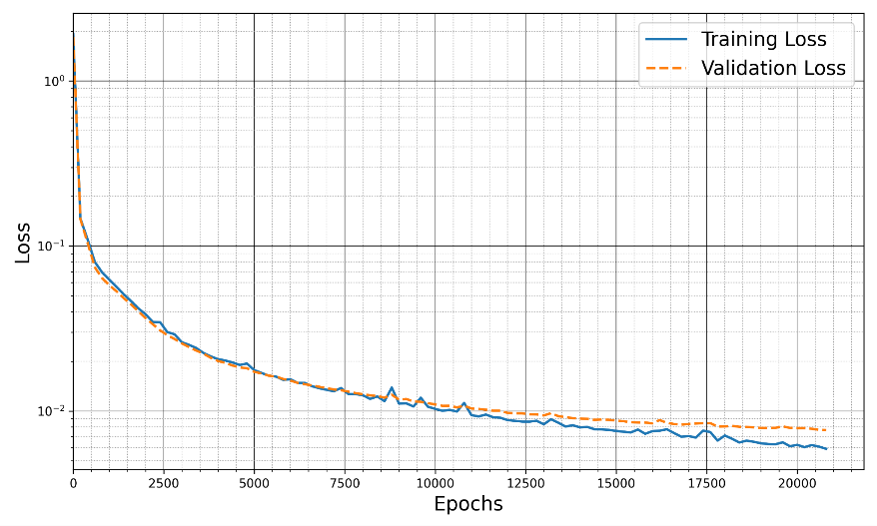}
\\[0.0em]
\textbf{Figure~\thefigure:} Training and validation loss histories of tLaSDI-pGFINN. 
\label{fig:Figure28}
\end{center}

The prediction performance comparison between the two models is summarized in Figure~\ref{fig:Figure29} and Table~\ref{tab:test_metrics3}. It can be observed that tLaSDI-pGFINN consistently outperforms AE-LSTM on unseen testing samples, as indicated by the lower mean squared error (MSE) and relative L2 errors.
\begin{center}
\refstepcounter{figure}
\includegraphics[width=0.85\textwidth]{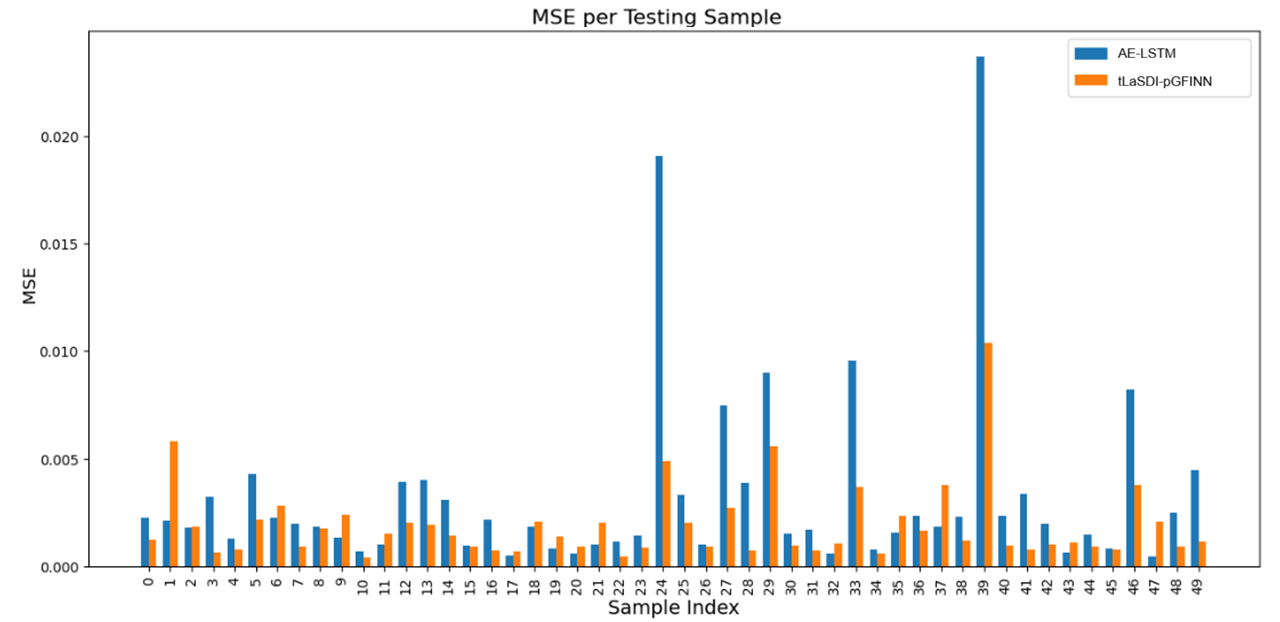}
\\[0.0em]
\textbf{Figure~\thefigure:} Comparison of MSE for each testing sample between AE-LSTM and tLaSDI-pGFINN.
\label{fig:Figure29}
\end{center}

\begin{table}[!htbp]
\centering
\refstepcounter{table}
\label{tab:test_metrics3}

{\bfseries Table \thetable. Prediction performance on the Testing Dataset}

\vspace{0.3em}

\begin{tabular}{lcc}
\hline
& \multicolumn{2}{c}{Model} \\
\cline{2-3}
Metric & AE-LSTM & tLaSDI-pGFINN \\
& w/ GC & w/GC \\
\hline
MSE & $3.23\times10^{-3}$ & $1.89\times10^{-3}$ \\
Average L2 Error & 12.21\% & 10.40\% \\
\hline
\end{tabular}

\end{table}

Figure~\ref{fig:Figure30} to Figure~\ref{fig:Figure32} show predictions with initial cracks located close to one of the inclusion. In Figure~\ref{fig:Figure30}, tLaSDI-pGFINN accurately captures the complete crack evolution, whereas AE-LSTM fails to predict the correct crack pattern and does not capture the fully developed crack at the final time step. Similarly, in Figure~\ref{fig:Figure31} and Figure~\ref{fig:Figure32} where the crack propagates around the inclusion, tLaSDI-pGFINN demonstrates better prediction performance compared to AE-LSTM, which produces inaccurate crack patterns and higher MSE values. In cases where the initial crack is located relatively far from the inclusions, the crack patterns are straighter as shown in Figure~\ref{fig:Figure33} and Figure~\ref{fig:Figure34}. The tLaSDI-pGFINN model still produces accurate predictions, while AE-LSTM model fails to capture the fully developed crack evolution, resulting in higher MSE values at the final time step. The superior performance of tLaSDI-pGFINN can be attributed to its thermodynamically consistent latent dynamics framework, which enforces the underlying physical constraints throughout the crack evolution process. This behavior is further demonstrated in Figure~\ref{fig:Figure35}, where the thermodynamic constraints remain satisfied throughout the time histories. The entropy plots exhibit a continuous increase over time, which is associated with the irreversible fracture processes. These observations further highlight the ability of the tLaSDI-pGFINN framework to capture physically consistent crack evolution while preserving the underlying thermodynamic principles.

\begin{center}
\refstepcounter{figure}
\includegraphics[width=0.65\textwidth]{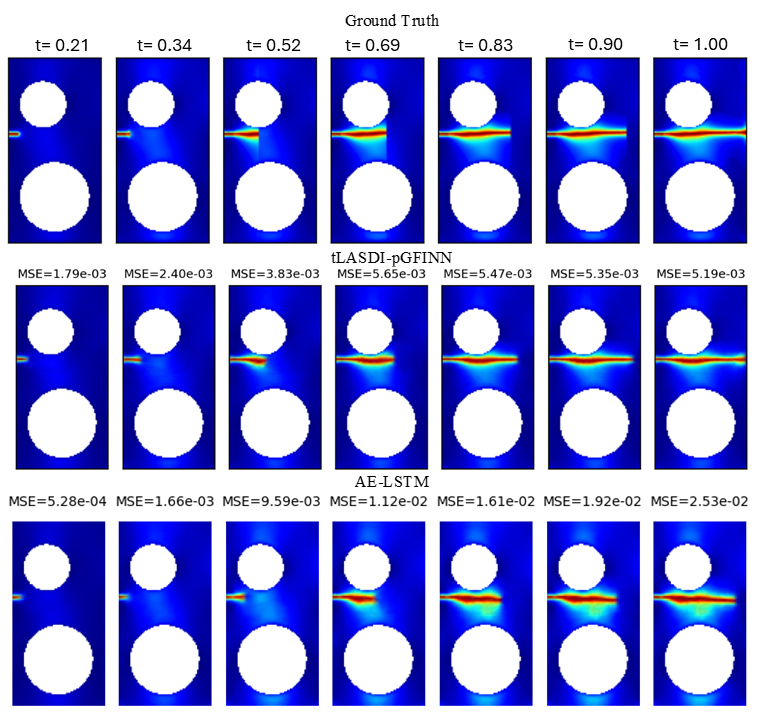}
\\[0.0em]
\textbf{Figure~\thefigure:} tLaSDI-pGFINN vs AE-LSTM prediction for testing sample 33.
\label{fig:Figure30}
\end{center}

\begin{center}
\refstepcounter{figure}
\includegraphics[width=0.65\textwidth]{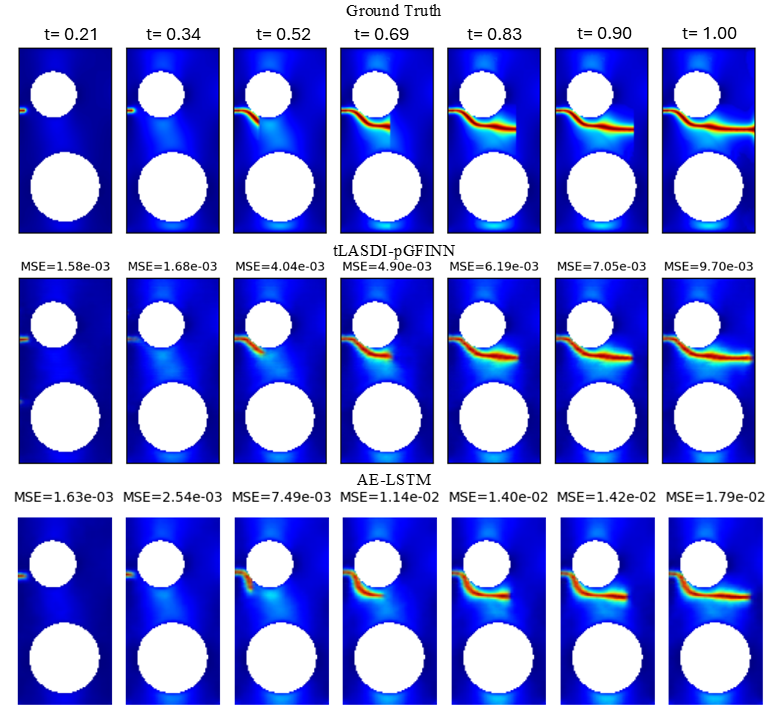}
\\[0.0em]
\textbf{Figure~\thefigure:} tLaSDI-pGFINN vs AE-LSTM prediction for testing sample 46.
\label{fig:Figure31}
\end{center}

\begin{center}
\refstepcounter{figure}
\includegraphics[width=0.65\textwidth]{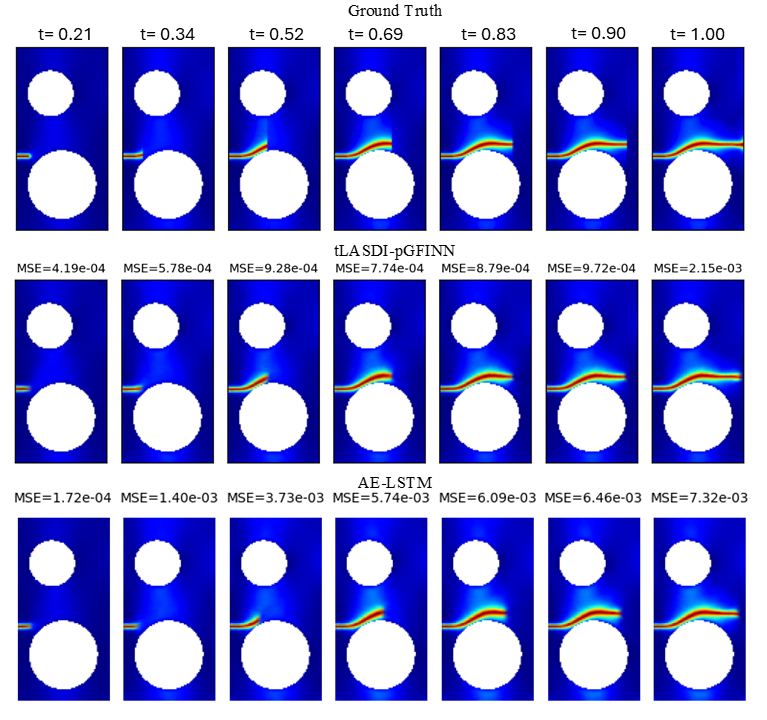}
\\[0.0em]
\textbf{Figure~\thefigure:} tLaSDI-pGFINN vs AE-LSTM prediction for testing sample 28.
\label{fig:Figure32}
\end{center}

\begin{center}
\refstepcounter{figure}
\includegraphics[width=0.65\textwidth]{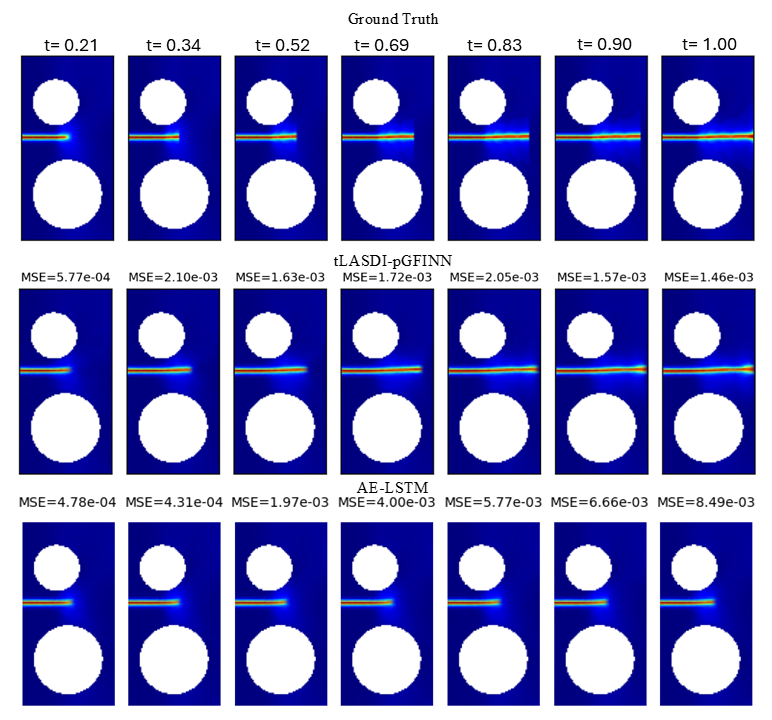}
\\[0.0em]
\textbf{Figure~\thefigure:} tLaSDI-pGFINN vs AE-LSTM prediction for testing sample 14.
\label{fig:Figure33}
\end{center}

\begin{center}
\refstepcounter{figure}
\includegraphics[width=0.65\textwidth]{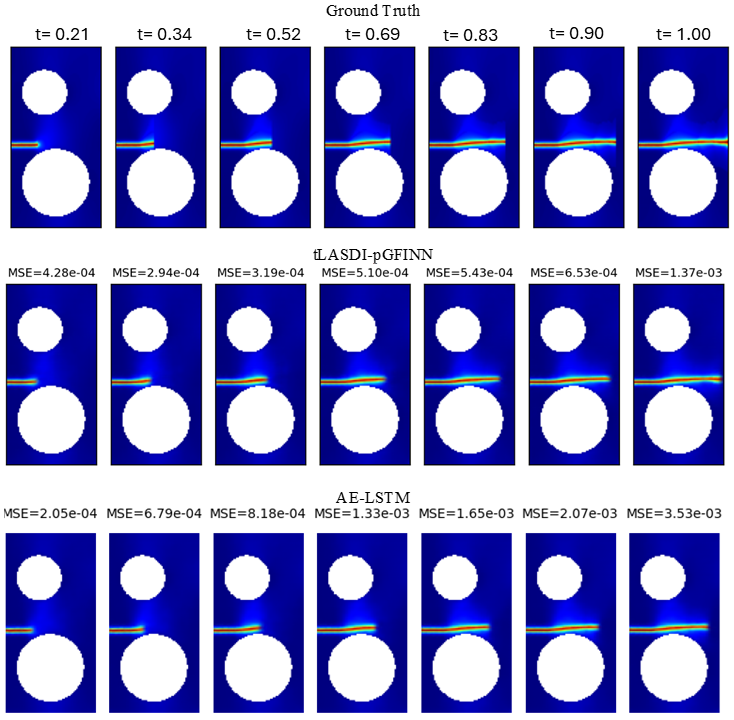}
\\[0.0em]
\textbf{Figure~\thefigure:} tLaSDI-pGFINN vs AE-LSTM prediction for testing sample 22.
\label{fig:Figure34}
\end{center}

\begin{center}
\refstepcounter{figure}
\includegraphics[width=0.95\textwidth]{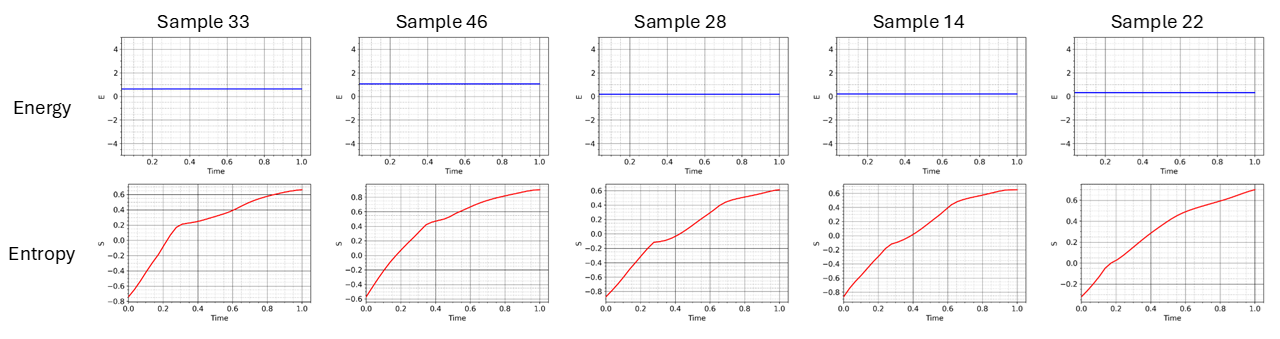}
\\[0.0em]
\textbf{Figure~\thefigure:} Energy and entropy histories predicted by pGFINN.
\label{fig:Figure35}
\end{center}

Finally, it is concluded that since the crack paths are strongly affected by their interactions with the inclusions, the geometry conditioner provides spatial and interface information to improve crack reconstruction. Table~\ref{tab:timing3} summarizes the training and prediction times of both models. The incorporation of the geometry autoencoder architecture and the larger spatial dimensions of each sample increase the training time for both models. Nevertheless, once trained, tLaSDI-pGFINN achieves higher accuracy than AE-LSTM in fracture prediction, while requiring only approximately 7 seconds per test case.

\begin{table}[!htbp]
\centering
\refstepcounter{table}
\label{tab:timing3}

{\bfseries Table \thetable. Training and Prediction Times of AE-LSTM and tLaSDI-pGFINN}

\vspace{0.3em}

\begin{tabular}{lcc}
\hline
Model & Training & Prediction \\
      & Time & Time \\
\hline
AE-LSTM & $\sim$35 mins & $\sim$5 s \\
tLaSDI-pGFINN & $\sim$11 Hours & $\sim$7 s \\
\hline
\end{tabular}

\end{table}

\section{Conclusions}
This study proposes tLaSDI-pGFINN for predicting crack propagation in heterogeneous materials. The proposed framework incorporates the GENERIC formalism into the neural network architecture to enforce thermodynamic consistency in the latent-space representation of the damage field. This is achieved by learning operator matrices that satisfy the symmetry properties and degeneracy conditions required by the GENERIC framework. To further improve the reconstruction of fracture evolution, a Geometry-Conditioned Autoencoder is introduced to more accurately capture material interface features governing crack initiation and propagation. With thermodynamics constraints inherently satisfied by construction, the model is trained using a comprehensive loss function designed to ensure accurate dynamic mappings between the latent and physical spaces.

Three numerical examples with varying levels of complexity are studied to investigate the performance of the proposed model: (1) a single-cavity problem with varying loading conditions, (2) a single-inclusion problem with varying inclusion locations and diameters, and (3) a multiple-inclusion problem with varying crack initiation locations. In validation and testing, the prediction performance of the model is evaluated using an autoregressive rollout strategy, allowing the true generalization capability of the framework to be assessed. All three numerical examples show good agreement with the reference solutions, where tLaSDI-pGFINN predicts fracture evolution more accurately than AE-LSTM in approximately 70\%–75\% of the testing samples. This indicates that the proposed framework not only learns the data distribution, but also captures the underlying physical behavior governing fracture evolution. The crack evolution plots demonstrate that tLaSDI-pGFINN can predict crack patterns that closely match the ground truth solutions. The proposed framework also outperforms AE-LSTM in cases involving complex crack topologies and delayed crack initiation, which are particularly challenging for purely data-driven models that rely only on temporal information in the recurrent cell state. The strong generalization capability of tLaSDI-pGFINN is further supported by the energy and entropy histories obtained from unseen testing samples, where both conservation and dissipation laws remain satisfied throughout the prediction process. Overall, this study demonstrates that incorporating thermodynamic constraints into the learning framework leads to more accurate, stable, and physically consistent predictions compared to models without the embedded physical principles.

\section*{Acknowledgements}
This work was supported by the Department of Energy National Nuclear Security Administration (DOE/NNSA) PSAAP IV program under Award Number DE-NA0004262 and the San Diego Supercomputer Center (SDSC) to the University of California San Diego, USA, and Ministry of Education (MOE), Taiwan, Yushan Fellow program under project contract number MOE-113-YSFEE-0005-001-P1 to National Cheng Kung University, Taiwan. The financial support by the Indonesia Endowment Fund for Education (LPDP) under the Ministry of Finance of the Republic of Indonesia to the first author is also greatly acknowledged.
\FloatBarrier

\bibliographystyle{cas-model2-names}
\bibliography{references}
\end{document}